\documentclass[pre,twocolumn,showpacs]{revtex4}
\usepackage{graphicx}
\usepackage{amsmath}
\usepackage{ifthen}

\newcommand{\ColBW}{color}

\newcommand{\md}[0]{\mbox{d}}

\begin{document}

\title{Coefficient of tangential restitution for the linear dashpot model}
  
\author{Volker Becker}
\affiliation{Charit\'e, Augustenburger Platz 1, D-13353 Berlin, Germany}
\author{Thomas Schwager}
\affiliation{Charit\'e, Augustenburger Platz 1, D-13353 Berlin, Germany}
\author{Thorsten P\"oschel}
\affiliation{Physikalisches Institut, Universit\"at Bayreuth, D-95440 Bayreuth, Germany}

\date{\today}

\begin{abstract}
  The linear dashpot model for the inelastic normal force between
  colliding spheres leads to a constant coefficient of normal
  restitution, $\varepsilon_n=$const., which makes this model very
  popular for the investigation of dilute and moderately dense
  granular systems. For two frequently used models for the
  tangential interaction force we determine the coefficient of
 tangential  restitution $\varepsilon_t$, both analytically and by
  numerical integration of Newton's equation. Although
  $\varepsilon_n=$const. for the linear-dashpot model, we obtain
  pronounced and characteristic dependencies of the tangential
  coefficient on the impact velocity
  $\varepsilon_t=\varepsilon_t\left(\vec{g}\right)$. The results may
  be used for event-driven simulations of granular systems of
  frictional particles.
  \end{abstract}


\maketitle

\section{Introduction}
The interaction of dissipatively colliding spheres is described by the interaction force law. In the case of smooth spheres, that is nonfrictional spheres, the force is directed along the  unit vector $\vec{e}\equiv \left(\vec{r}_i-\vec{r}_j\right)/\left|\vec{r}_i-\vec{r}_j\right|$. Thus, only the normal component of the impact velocity, $g_n\equiv \left(\vec{v}_i-\vec{v}_j\right) \cdot \vec{e}$, (with $\vec{v}_{i/j}$ being the center-of-mass velocities of the particles $i$ and $j$) changes during the contact. The normal component of the velocity after a collision is then obtained by integrating Newton's equation of motion for the mutual deformation $\xi$ of the particles
\begin{equation}
  \label{eq:NewtonNormal}
  m_\text{eff}\ddot{\xi}+F_n\left(\dot{\xi},\xi\right) = 0\,,~~~\dot{\xi}(0)=g_n\,,~~~\xi(0)=0\,,
\end{equation}
with $\xi(t)\equiv \max\left(0, R_i+R_j-\left|\vec{r}_i-\vec{r}_j\right|\right)$ and where $F_n$ is the model-specific interaction law.

Alternatively, the collision may be described using the coefficient of restitution, relating the pre-collisional and post-collisional normal relative velocities,
\begin{equation}
  \label{eq:eps_n_def}
  \varepsilon_n\equiv - \frac{g_n^\prime}{g_n}=-\frac{\dot{\xi}\left(t_c\right)}{\dot{\xi}(0)}
\end{equation}
with $t_c$ being the duration of the collision. Throughout this paper, primed variables denote post-collisional quantities. Consequently, the choice of the particular force $F_n$ determines the coefficient of restitution $\varepsilon_n$. 

From its definition, Eq. \eqref{eq:eps_n_def}, obviously $0\le \varepsilon_n \le 1$ where $\varepsilon_n=1$ describes an elastic collision. The assumption $\varepsilon_n=$const. is widely used throughout the literature on granular gases and other dilute granular systems. This assumption is not in perfect agreement with physical reality, see \cite{Ramirez:1999}, but it simplifies the analysis of kinetic and hydrodynamic equations considerably and its use is, therefore, justified.   

Vice versa, one may ask which force laws lead to $\varepsilon_n=$const. For the rather general Ansatz $F_n = F_n^\text{(el)} + F_n^\text{(dis)}$ with $F_n^\text{(el)}\propto \xi^a$, $F_n^\text{(dis)}\propto \dot{\xi}^b\xi^c$ for the elastic and dissipative components of the interaction force, a dimension analysis \cite{Taguchi:1992JDP,LudingClementBlumenRajchenbachDuran:1994DISS,Ramirez:1999} shows that only combinations with $2(c-a)+b(1+a)=0$ lead to $\varepsilon_n=$const. Assuming $b=1$, i.e., a linear dependence of the dissipative force on the velocity (which should  be justified at least for small impact rate), we see that the choice $a=1$, $c=0$ fulfills the above condition. This corresponds to the linear dashpot model
\begin{equation}
  \label{eq:dashpot}
  F_n=-k_n\xi - \gamma_n\dot{\xi}\,.
\end{equation}
Indeed, the linear dashpot model is frequently used in Molecular Dynamics simulations of granular systems, e.g. \cite{Taguchi:1992JDP,GallasHerrmannSokolowski:1992PRL,LandryGreast:2004PRE,KohringMelinPuhlTillemans:1994,TsimringVolfson:2005,SilbertErtasGrestHalseyLevine:2002}.
The coefficient of restitution can be found by integrating Eq. \eqref{eq:NewtonNormal} with the definition Eq. \eqref{eq:eps_n_def}, where the end of the collision $t_c>0$ is given by the condition $\ddot{\xi}=0$ which takes into account that there is only repulsive interaction between granular particles, see \cite{SchwagerPoeschel:2007} for details. The coefficients of the force law, $k_n$ and $\gamma_n$ translate into the impact-velocity independent coefficient of restitution \cite{SchwagerPoeschel:2007}
\begin{equation}
  \varepsilon_n = \begin{cases}
    \displaystyle\exp\left[-\frac{\beta}{\omega_n}\left(\pi - \arctan\displaystyle\frac{2\beta\omega_n}{\omega_n^2-\beta^2}\right)\right],~~~  \displaystyle\beta<\frac{\omega_0}{\sqrt{2}} \\[0.3cm]
    \displaystyle\exp\left[-\frac{\beta}{\omega_n}\arctan\displaystyle\frac{2\beta\omega_n}{\omega_n^2-\beta^2}\right],
    ~~~ \displaystyle \beta\in\left[\frac{\omega_0}{\sqrt{2}},\omega_0\right]
\\[0.3cm]
    \displaystyle\exp\left[-\frac{\beta}{\omega_n}\ln\frac{\beta+\omega_n}{\beta-\omega_n}\right],~~~ \beta>\omega_0
    \end{cases}
  \label{eq:dashpotCOR1}
\end{equation}
\begin{equation}
  \omega_0^2\equiv\frac{k_n}{m_\text{eff}}\,;~\beta\equiv
  \frac{\gamma_n}{2m_\text{eff}}\,;~\omega_n^2\equiv\omega_0^2-\beta^2\,;~\omega_n^2 = \beta^2-\omega_0^2\,.
  \label{eq:omegadef}
\end{equation}

There are several other force laws in the literature, for review see \cite{SchaeferDippelWolf:1995,KruggelEtAl:2007}. Some of them are certainly better suited to describe the mechanics of colliding spheres, however, none of them leads to a constant coefficient of normal restitution. The condition $\varepsilon_n=$const. in turn is essential for an entire class of scientific literature in the field of dilute granular gases. Therefore, here we restrict ourselves to this important case.

For the case of frictional particles, in general, a particle-particle interaction causes not only a change in the normal component of the relative velocity  but also a change of its tangential component as well as the particles' rotational velocity. Let us denote the relative velocity of the particles {\em in the point of contact} by
\begin{equation}
  \label{eq:g_def}
  \vec{g}\equiv \vec{v}_i-\vec{v}_j-\left(R_1\vec{\Omega}_1+R_2\vec{\Omega}_2\right)\times\vec{e} 
\end{equation}
with $\vec{\Omega}_{i/j}$ and $R_{i/j}$ being the angular velocities and radii of the two particles. Its projection to the tangential plane in the point of contact reads
\begin{equation}
  \label{eq:g_tan_def}
  \vec{g}_t\equiv -\vec{e}\times\left(\vec{e}\times\vec{g}\,\right)\,.
\end{equation}
Similar to the normal direction, the change of the velocity in tangential direction is described by the coefficient of tangential restitution, 
\begin{equation}
  \label{eq:eps_t_def}
  \varepsilon_t\equiv \frac{g_t^\prime}{g_t}\,.
\end{equation}
In contrast to the coefficient of normal restitution, here $-1\le \varepsilon_t\le 1$, that is, there are two elastic limits. The case $\varepsilon_t=1$ corresponds to smooth particles, that is, the tangential velocity and, thus, the angular velocities of colliding particles do not change. The other elastic case $\varepsilon_t=-1$ corresponds to rough particles. One may think (in 2d) of gear wheels made of a very elastic material. When such particles collide, the tangential component of their relative velocity is reverted. The case, $\varepsilon_t=0$, describes the total loss of relative tangential velocity after a collision.

While the normal force is given by Eq. \eqref{eq:dashpot}, commonly in Molecular Dynamics simulations the change of the tangential velocity during an impact is described by tangential force laws which will be introduced in the next section. In a similar way as described above for the coefficient of normal restitution, one can analyze the tangential relative velocity of colliding particles to obtain the coefficient of tangential restitution $\varepsilon_t$. As the coefficients of restitution are a direct consequence of the actual trajectory of the particles during contact, its functional dependence on the material properties and the impact velocities depends on the chosen force law. 

It is the aim of this paper to characterize the coefficient of tangential  restitution for different expressions for the tangential interaction force between colliding particles as commonly used in Molecular Dynamics simulation. In particular, we are interested in the important case that the coefficient of normal restitution is independent of the impact velocity, $\varepsilon_n=\mbox{const.}$ as it follows from the linear dashpot model, Eq. \eqref{eq:dashpot}.

\section{Tangential Forces}

\subsection{Coulomb law for static friction}

The normal force between contacting spheres is determined by their mutual compression $\xi$ and the compression rate $\dot{\xi}$. This is true not only for the linear dashpot model, Eq. \eqref{eq:dashpot}, but more generally for all non-adhesive collision models, see \cite{SchaeferDippelWolf:1995,KruggelEtAl:2007}, and is due to the fact that the interaction force is a bulk material property. In contrast, the tangential force is not only a bulk property but also significantly determined by surface properties, e.g. roughness. 

The usual textbook formulation of friction distinguishes between static and dynamic friction. If the particles in contact slide on each other the friction force $F_t$ (absolute value) is
\begin{equation}
  F_t = \mu F_N\,,
  \label{eq:coulomblimit}
\end{equation}
where $F_N$ is the absolute value of the normal force at contact. Thus, the tangential force is limited by the Coulomb friction law, Eq. \eqref{eq:coulomblimit}. If the particles do not slide on each other, i.e. if the tangential relative velocity at contact is zero ($g_t=0$), the friction force is only indirectly defined. Namely, it assumes the value necessary to keep the particles from sliding as long as the resulting force does not exceed the Coulomb limit, Eq. \eqref{eq:coulomblimit}. Hence, in this formulation there is no force law for static friction -- the friction force is not determined from geometric properties like deformation. This makes the application of the Coulomb friction law in Molecular Dynamics simulations difficult: For the numerical integration of Newton's equation we need in each time step the forces acting on the particles. These forces must be expressed in terms of positions and velocities of the particles. Therefore, Coulomb's law which can \`a priori not be expressed in terms of positions and velocities must be modeled by a function in these variables,
\begin{equation}
  \label{eq:F_t_general}
  F_t=-\min\begin{cases} 
    \mu F_n\\
    f\left(\vec{v}_1, \vec{v}_2, \dot{\vec{v}}_1, \dot{\vec{v}}_2,\dots\right)\,,
  \end{cases}
\end{equation}
where the model specific functional $f()$ depends on the history of its arguments, $\vec{v}_1(\tau)$; $0\le\tau\le t$, and the other arguments likewise. The choice of this functional is not unique but ambiguous to a certain degree. 

\subsection{Model by Haff and Werner}
The most simple representation of the force scheme Eq. \eqref{eq:F_t_general} is the model by Haff and Werner \cite{HaffWerner:1986}
\label{sec:Haff}
\begin{equation}
  \label{eq:F_t_Haff}
  F_t=-\min\left[\mu F_n, \gamma_t g_t(t)\right]\,,
\end{equation}
with the tangential component of the relative velocity at the point of contact, $g_t(t)$, given in Eq. \eqref{eq:g_def}. (Throughout this paper we call the components of the relative velocity {\em before} the collision $g_n$ and $g_t$. Time dependent velocities that vary {\em during} the collision are called as $g_n(t)$ and $g_t(t)$. The final velocity components are named $g_n^\prime$ and $g_t^\prime$.) 
Without loss of generality, here and in the following the tangential velocity at the contact point shall be positive. The case of negative tangential velocity can be deduced by reflection and leads to identical results.

Thus, the model by Haff and Werner assumes shear damping $\sim g_t$ for small velocity, limited by Coulomb's law. The model was successfully applied in many Molecular Dynamics simulations of granular matter although there appear problems when rather static systems are simulated, see \cite{PoeschelSchwager:2005} for a detailed discussion. 

However, if we consider the coefficient of normal restitution which corresponds to the model, Eq. \eqref{eq:F_t_Haff}, we notice more serious difficulties: both alternatively acting tangential forces due to Eq. \eqref{eq:F_t_Haff} cannot lead to negative relative velocities, $g_t$, after a collision. This is obvious for $F_t\propto -g_t$ and will be shown in Sec. \ref{sec:PureCoulomb} for $F\propto -F_n$. Consequently, the model by Haff and Werner cannot yield coefficients of tangential restitution of negative value. This is a serious inadequacy of this model. We will come back to this problem in Sec. \ref{sec:Haff_analyse}.

\subsection{Model by Cundall and Strack}
\label{sec:CundallEps}

The model by Cundall and Strack \cite{CundallStrack:1979} mimics static friction by means of a spring acting in tangential direction with respect to the contact plane. The spring is initialized at the time of first contact, $t=0$, and exists until the surfaces of the particles separate from one another after the collision. The elongation
\begin{equation}
  \label{eq:CundallSpring}
  \zeta(t)=\int\limits_0^t g^t\left(t^\prime\right)\md t^\prime
\end{equation}
characterizes the restoring tangential force, limited by Coulomb's law. Thus
\begin{equation}
  \label{eq:CundallForce}
  F_t=-\min\left(\mu F_n, k_t \zeta\right)\,,
\end{equation}
Consequently, when $\mu F_n < k_t \zeta$, that is, the Coulomb law applies, the spring assumes the elongation $\zeta=\mu F_n/k_t$. The energy stored in the spring may be released in a later stage of the collision.
From Molecular Dynamics simulation we know that this model is much better suited to describe static behavior of granular matter \cite{PoeschelSchwager:2005}. As shown below, it yields also negative values of the coefficient of tangential restitution for appropriate choice of the initial relative velocity at the point of contact, e.g., $g_n$ and $g_t$. This is due to the fact that the internal spring acts like a reservoir of energy for the relative motion of the particles in tangential direction. In the first part of the collision the spring is loaded and may release the stored energy towards the end of the collision. This way, the tangential component of the relative velocity may change its sign. The coefficient of tangential restitution which corresponds to the model by Cundall and Strack, Eq. \eqref{eq:CundallForce} is discussed in detail in Sec. \ref{sec:CundallEps}.

\section{Coefficient of tangential restitution}
\label{sec:eps}

\subsection{Pure Coulomb Force}
\label{sec:PureCoulomb}

Before discussing the most common tangential forces used in Molecular Dynamics simulations let us derive some general expressions which apply to the limiting case of pure sliding: Thus, for the moment we assume simplifying that during the entire collision the friction force is not sufficient enough to stop the tangential relative motion. Consequently, $F_t=\mu F_n$, from the beginning to the end of the collision. The results derived here are valid independently of the functional form of the normal force. Therefore, the function $F_n$ remains unspecified, except for the fact that it is a function of time, defined in the interval $\left(0,t_c\right)$. 

During the contact the change of the velocities in normal and tangential direction obey Newton's law,
\begin{equation}
  \label{eq:eq_mot}
  \frac{d g_n(t)}{dt} = \frac{1}{m_\text{eff}} F_n\,;~~~~~
  \frac{d g_t(t)}{dt} = \frac{1}{\alpha} F_t 
\end{equation}
with
\begin{equation}
  \alpha\equiv\left[\frac{1}{m_\text{eff}}+\frac{R_1^2}{J_1} + \frac{R_2^2}{J_2} \right]^{-1}\, .
  \label{eq:alphadef}
\end{equation}
which may be obtained from Eq. \eqref{eq:g_tan_def} with the general paradigm of nearly instantaneous collisions, that is, the unit vector $\vec{e}$ does not change during the collision. The validity of this approximation will be discussed briefly in Sec. \ref{sec:unitvector}.

Using the definition of the coefficient of normal restitution, Eq. \eqref{eq:eps_n_def}, we write for the change of the normal component
\begin{equation}
  \int_0^{t_c}F_n(t)\md t = \left(1+\varepsilon_n\right)m_\text{eff}g_n\,.
\end{equation}
Again we assume the tangential velocity at contact to be positive.
During the collision the friction force assumes the value $F_t=-\mu
F_n(t)$, therefore, the tangential component of the relative velocity in the point of contact
after the collision reads
\begin{equation}
  \begin{split}
  g_t^\prime - g_t &= \frac{1}{\alpha}\int_0^{t_c} F_t\md t  = -\frac{\mu}{\alpha} \int_0^{t_c} F_n\md t\\
  &= -\frac{\mu\left(1+\varepsilon_n\right)m_\text{eff}}{\alpha}g_n\\
  \end{split}
\end{equation}
Hence, with the definition of the coefficient of tangential restitution we find the relation
\begin{equation}
  \label{eq:coulomb_eps}
   \varepsilon_t = 1 -\frac{\mu\left(1+\varepsilon_n\right)m_\text{eff}}{\alpha}~ \frac{g_n}{g_t} \,,
\end{equation}
independent of the functional form of the normal and tangential force laws (see \cite{WaltonBraun:1986,Luding:1995}).

As an important consequence we find that 
the coefficient of tangential restitution does significantly depend on both the normal and the tangential relative velocities. Formally, Eq. \eqref{eq:coulomb_eps} leads to values of $\varepsilon_t$ outside its range of definition. However, the basic assumption of pure Coulomb friction implies that the particles do not stop sliding on each other, that is, during the entire contact the particle remain in the Coulomb regime where $F_t=-\mu F_n$. Hence $g_t^\prime\ge 0$ and, thus
$\varepsilon_t\ge 0$.

\begin{figure}
  \centering
  \ifthenelse{\equal{\ColBW}{color}}{
  \includegraphics[angle=0,bb=32 26 378 278,clip,width=0.99\columnwidth]{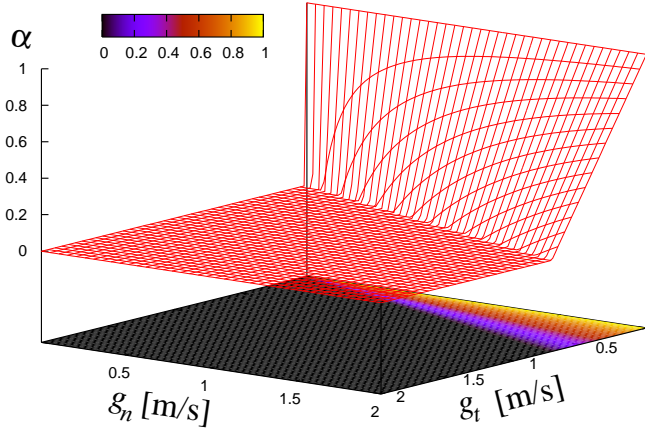}}
{
   \includegraphics[angle=0,bb=32 26 378
  278,clip,width=0.99\columnwidth]{figs/purecoloumb/purecoloumb.eps}}
  \caption{The coefficient of tangential restitution as function of
    the tangential impact velocity $g_t$ and the normal impact
    velocity $g_n$ in the case of pure Coulomb friction. The
    parameters are $\mu=0.4$ and $k_n=10^6$~N/m.}
  \label{fig:eps_coulomb}
\end{figure}

Consequently, for pure Coulomb friction we obtain
\begin{equation}
   \label{eq:coulomb_eps1}
    \varepsilon_t = \max \left(0, 1 -\frac{\mu\left(1+\varepsilon_n\right)m_\text{eff}}{\alpha}~ \frac{g_n}{g_t} \right)\,,
\end{equation}
see Fig. \ref{fig:eps_coulomb}. 

\subsection{Model by Haff and Werner}
\label{sec:Haff_analyse}
\label{sec:HaffResults}
\begin{figure}[htbp]
\centering
\includegraphics[width=0.9\columnwidth,clip]{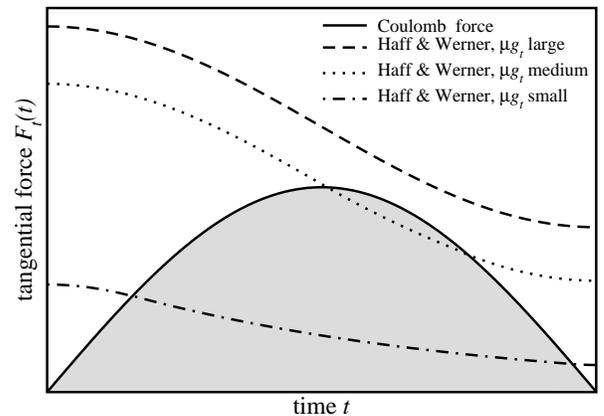}
\caption{Sketch of the tangential force $F_t(t)$. When the collision starts, the Coulomb force is always active since for finite $g_t$ always $\gamma_t g_t > \mu F_n = 0$. For large initial tangential velocity, the condition $\gamma_t g_t(t) < \mu F_n(t)$ is never fulfilled during the entire condition. For smaller $g_t$ the tangential velocity drops up to $t=t_{1,s}$ where $\gamma_t g_t(t_c) = \mu F_n(t_c)$ and the Haff-Werner force $\gamma_t g_t$ becomes active. The tangential velocity decays further and eventually at time $t_{s,2}$ the Coulomb force becomes active again. 
}
\label{fig:postimeev}
\end{figure}

First we want to present an analytic theory for the elastic collision
in normal direction, that is $\gamma_n=0$ in Eq. \eqref{eq:dashpot}. The time dependent solution of the corresponding Newton's equation, Eq. \eqref{eq:NewtonNormal}, reads
\begin{equation}
  \xi(t)=\frac{g_n}{\omega_n} \sin \omega_n t\,,
\end{equation}
where $\omega_n=\sqrt{k/m_\text{eff}}$.  Note that the compression $\xi(t)$ is independent of the tangential relative motion of the particles. 
With Eqs. \eqref{eq:coulomblimit} and \eqref{eq:dashpot}, the limiting Coulomb force is, therefore,
\begin{equation}
 \mu F_n = \frac{\mu g_n k_n}{\omega_n} \sin \omega_n t.
\end{equation}
When the collision starts at time $t=0$ at finite tangential velocity $g_t$, the magnitude of the Coulomb force $\mu F_n$ is always smaller than the magnitude of the shear damping force, $\gamma_t g_t$, that is, the tangential force in the beginning of the collision is equal to the Coulomb force. We integrate Eq. \eqref{eq:F_t_Haff} and find that during this first stage of the collision, the relative tangential velocity decays as  
\begin{equation} \label{eq:vtcoulombhw1}
  g_t(t)= g_t + \frac{\mu g_n m_\text{eff}}{\alpha} \left( \cos \omega_n t -1 \right)
\end{equation}
If the magnitude of the shear damping force drops below the Coulomb force, the tangential force is governed by the other branch of the force law, $F_t=\gamma_t g_t$. The transition takes place at time $t_{s,1}$ when 
\begin{equation}
  \frac{\mu k_n g_n}{\omega_n}\sin(\omega_n t_{s,1}) = \gamma_t g_t(t_{s,1})\,.  
\end{equation}
Inserting Eq. \eqref{eq:vtcoulombhw1} we obtain the first switching time $t_{s,1}$ between the regimes, 
\begin{equation}
  \label{eq:hwswitch1}
  \begin{split}
    \sin\omega_nt_{s,1} &= \gamma_t\frac{\alpha\omega_n(\delta-1) + \sqrt{\alpha^2\omega_n^2 - (\delta^2-2\delta)\gamma_t^2}}{\alpha^2\omega_n^2 + \gamma_t^2}\\
    \delta &= \frac{\alpha g_t}{\mu m_\text{eff}g_n}
\end{split}
\end{equation}
If the initial tangential velocity fulfills the inequality
\begin{equation}
  \frac{g_t}{g_n} > \frac{\mu m_\text{eff}}{\alpha} + \sqrt{\frac{\mu^2m_\text{eff}^2}{\alpha^2} + \frac{\mu^2m_\text{eff}k_n}{\gamma_t^2}}
\end{equation}
Eq. \eqref{eq:hwswitch1} does not have a real solution and the Coulomb regime is active during the entire collision, (cf. Fig. \ref{fig:postimeev}, dashed line).
In this case the coefficient of tangential restitution reads
\begin{equation}
  \varepsilon_t=1-\frac{2 \mu m_\text{eff}}{\alpha} \frac{g_t}{g_n}
\end{equation} 
which coincides with Eq. \eqref{eq:coulomb_eps} for the case $\varepsilon_n=1$. 

For smaller initial tangential velocities Eq. \eqref{eq:hwswitch1} has a solution, that is, after the time $t_{s,1}$ the Haff-Werner branch of the tangential force becomes active and the tangential velocity decays exponentially:
\begin{equation}
  g_t(t)=g_t(t_{s,1}) \exp \left[-\frac{\gamma_t}{\alpha} \left(t-t_{s,1}\right) \right]
\end{equation}

\begin{figure}
  \centering
   \ifthenelse{\equal{\ColBW}{color}}{
  \includegraphics[angle=-90,bb=191 237 419 583,clip,width=0.99\columnwidth]{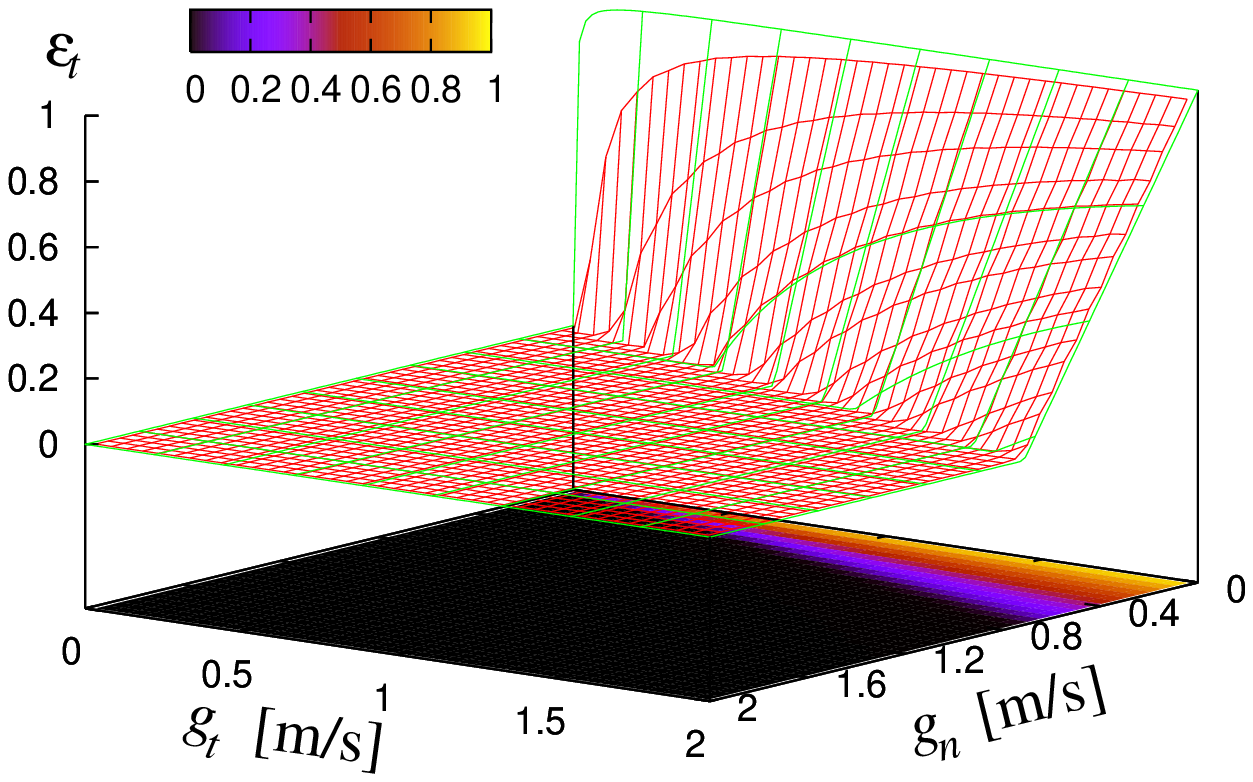}
  \includegraphics[angle=-90,bb=191 237 419 583,clip,width=0.99\columnwidth]{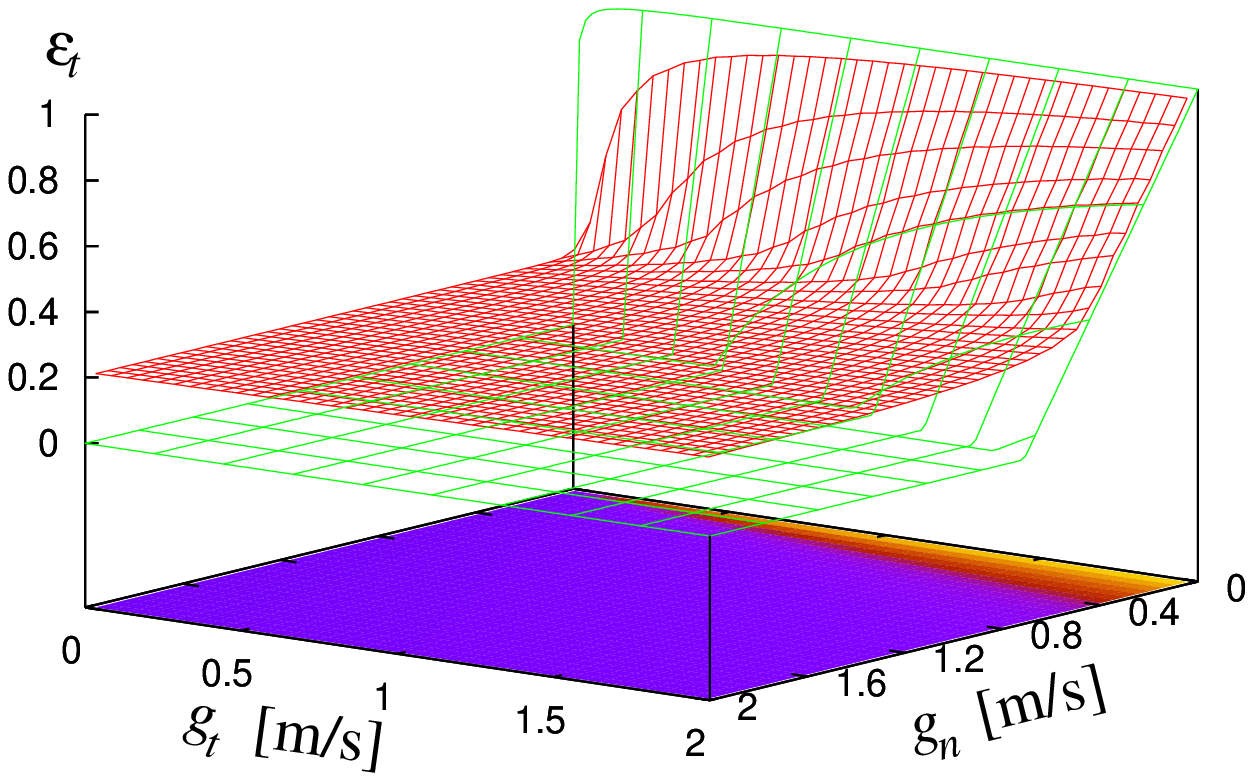}
  \includegraphics[angle=-90,bb=191 237 419 583,clip,width=0.99\columnwidth]{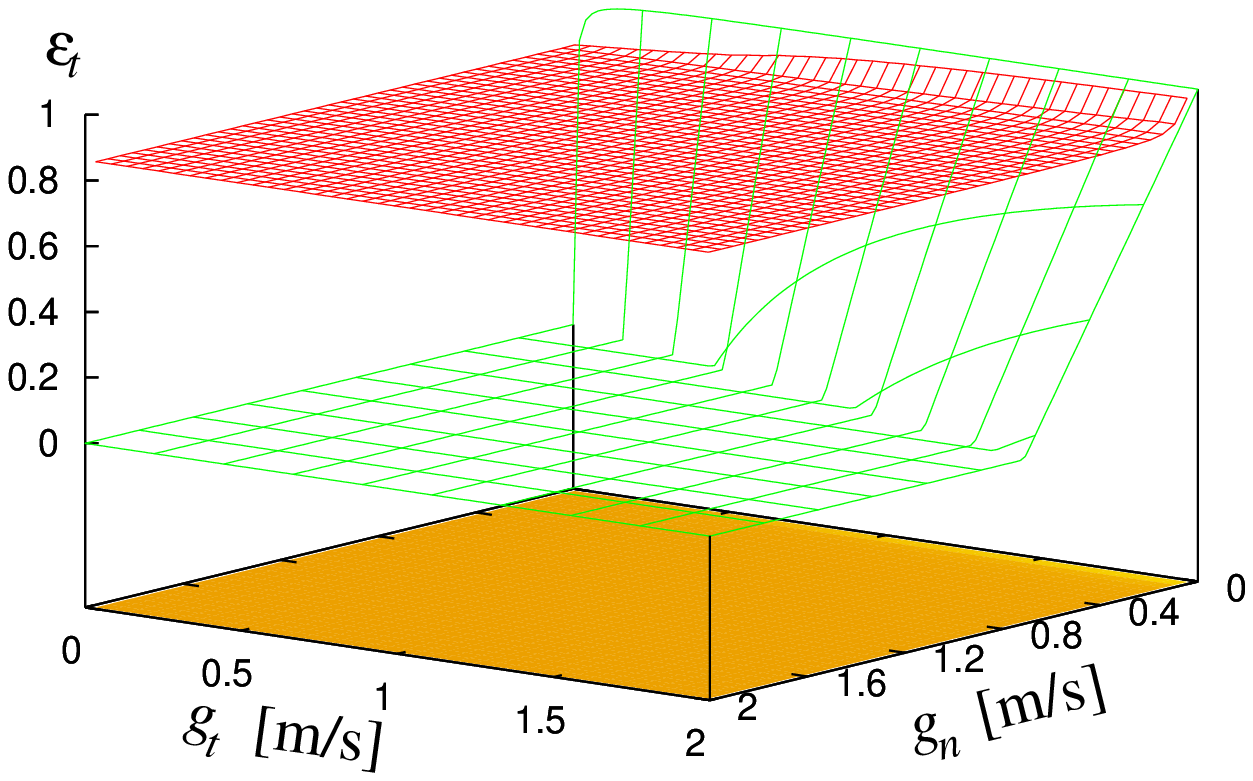}}
{
  \includegraphics[angle=-90,bb=191 237 419 583,clip,width=0.99\columnwidth]{figs/haffwerner/haffwerner3_pa.ps}
  \includegraphics[angle=-90,bb=191 237 419 583,clip,width=0.99\columnwidth]{figs/haffwerner/haffwerner2_pa.ps}
  \includegraphics[angle=-90,bb=191 237 419
  583,clip,width=0.99\columnwidth]{figs/haffwerner/haffwerner1_pa.ps}
}
  \caption{The coefficient of tangential restitution resulting from
    the model by Haff and Werner as function of the normal and
    tangential components of the impact velocity. The coarse lattices
    display the solution for pure Coulomb force,
    Eq. \eqref{eq:coulomb_eps1}. The parameters are $k_n=10^5$\,N/m,
    $\mu=0.4$, $\gamma_t = 10$~Ns/m (top), $\gamma_t=1$~Ns/m (middle),
    $\gamma_t=0.1$~Ns/m (bottom)}
  \label{fig:eps_haffwerner}
\end{figure}

Since the tangential velocity in the Haff-Werner model cannot drop to zero, at some later time $t_{s,2}>t_{s,1}$ the switching condition must be fulfilled again
\begin{equation}
  \frac{\mu g_nk_n}{\omega_n} \sin(\omega_n t_{s,2}) = \gamma_t g_t(t_{s,2})  \,,
\end{equation}
see Fig. \ref{fig:postimeev} (dotted and dash-dotted lines). This time is determined by 
\begin{equation}
  \exp \left[\frac{\gamma_t}{\alpha} \left(t_{s,2}-t_{s,1}\right)
  \right] \sin\left(\omega_n t_{s,2}\right) = g_t\left(t_{s,1}\right) \frac{\gamma_t\omega_n}{\mu k_n g_n}\,.
\end{equation}
For $t>t_{s,2}$ the Coulomb branch of the tangential force becomes active again for the rest of the collision. The final velocity reads
\begin{equation}
  g_t^\prime=g_t\left(t_{s,2}\right) - \frac{\mu g_n m_\text{eff}}{\alpha} \left[ \cos \left(\omega_n t_{s,2}\right)+1\right]
\end{equation}
and the coefficient of tangential restitution is 
\begin{equation}
  \varepsilon_t=\frac{g_t^\prime}{g_t}
\end{equation}
In the case of small tangential velocities, the time at the beginning
and the end of the collision where the Coulomb
regime is active is negligible and the decay of the tangential
velocity is mostly determined by the shear damping force. In this case the coefficient of restitution reads
\begin{equation} \label{eq:hw_small_gt}
  \varepsilon_t=\exp \left(-\frac{\pi \gamma_t }{\alpha \omega_n} \right)
\end{equation}
Figure \ref{fig:eps_haffwerner} shows the coefficient of tangential restitution as function of the normal and tangential components of the impact velocity. We determined the coefficient of tangential restitution also by solving the equation of motion, Eq. \eqref{eq:eq_mot},  numerically and analytically as described in this section, leading to perfect agreement. 

In the case of large tangential velocity $g_n$, the coefficient of restitution behaves like in the pure Coulomb case, that is, and for small values of $g_t$ the coefficient reaches the value predicted by 
Eq. \eqref{eq:hw_small_gt} (see Fig. \ref{fig:eps_haffwerner}, coarse lattices).

In a good approximation the coefficient of tangential  restitution in the model by Haff and Werner can conveniently be described as either a non-negative constant $\varepsilon_t^0$ or the dependence given by Eq. \eqref{eq:coulomb_eps}, whatever is larger:
\begin{equation}
  \varepsilon_t \approx \max\left(\varepsilon_t^0, 1- \frac{2\mu m_{\text{eff}}}{\alpha}\frac{g_n}{g_t}\right)\label{eq:epsHaffWernerAproximation}\,.
\end{equation}
Thus, the heuristics used in \cite{WaltonBraun:1986,FoersterLougeChangAllia:1994,Luding:1995} (Eq. \eqref{eq:epsHaffWernerAproximation} in a different notation) is justified, provided $\varepsilon_t^0\ge0$.

\subsection{Model by Cundall and Strack}
\label{sec:CundallResults}
\subsubsection{Equations of motion}
The collision model by Cundall and Strack is described by Eqs. (\ref{eq:NewtonNormal}, \ref{eq:eq_mot}) and the force laws, Eqs. (\ref{eq:dashpot}, \ref{eq:CundallForce}). The equation of motion reads, thus,
\begin{equation} 
  \label{eq:eq_mot_CS}
  \begin{split}
  m_\text{eff}\ddot{\xi} + k_n\xi + \gamma_n \dot \xi =& 0\\
  \dot{\zeta} =& g_t(t)\\
  \alpha\dot{g}_t - F_t(\zeta, F_N) =& 0    \\
  \xi(0) = 0\,;~~\dot{\xi}(0)= g_n\,;~~\zeta(0) = 0\,;~~\dot{\zeta}(0) &= g_t
  \end{split}
\end{equation}
The tangential force
\begin{equation}
  \left|F_t\right| = \min\left(k_t\zeta, \mu k_n\xi\right)
\end{equation}
counteracts a the elongation of the tangential spring, that is,
\begin{equation}
  F_t = -\mbox{sgn}\zeta\min\left(k_t\zeta, \mu k_n\xi\right)\,.
\end{equation}
The equation of motion for the tangential degree of freedom thus reads
\begin{equation} 
  \label{eq:eq_mot_tan_CS}
  \alpha\ddot{\zeta} + \mbox{sgn}\zeta\min\left(k_t\zeta, \mu k_n\xi\right) = 0\,.
\end{equation}

\subsubsection{Numerical results}
\begin{figure}[htbp]
  \centering
  \includegraphics[angle=0,bb=19 8 246 188,clip,width=0.99\columnwidth]{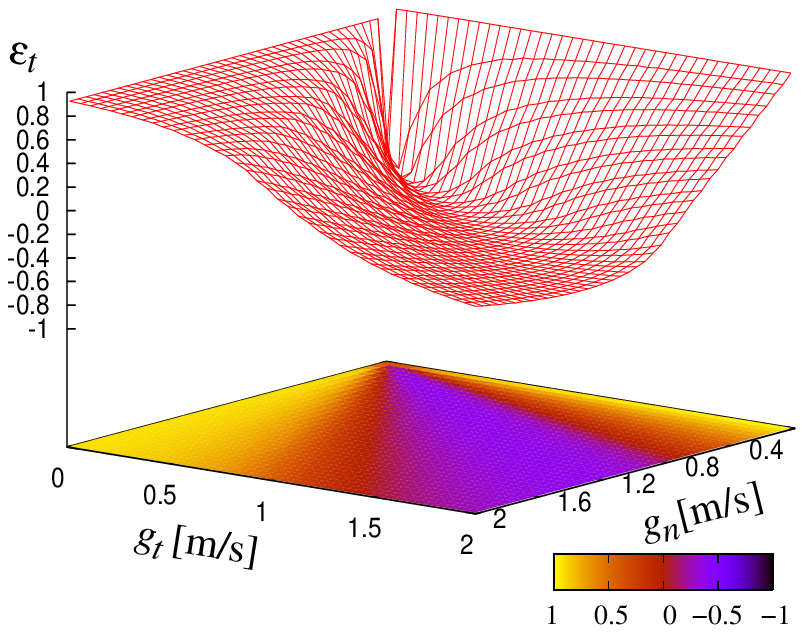} 
  \includegraphics[angle=0,bb=34 38 248 188,clip,width=0.49\columnwidth]{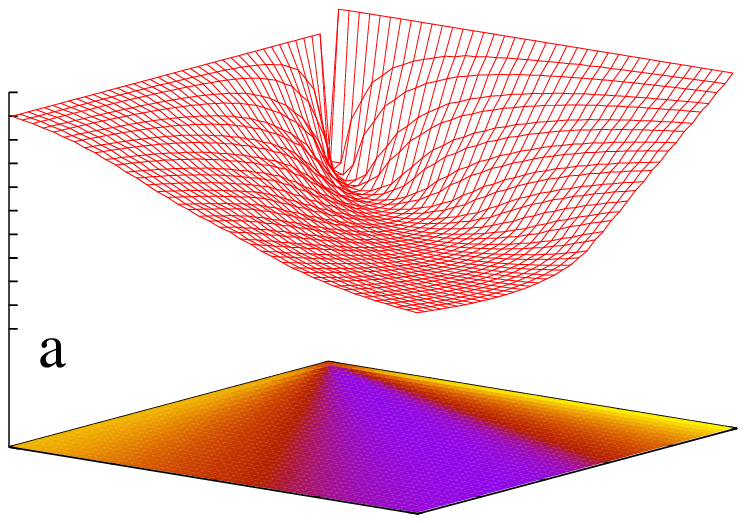}\includegraphics[angle=0,bb=34 38 248 188,clip,width=0.49\columnwidth]{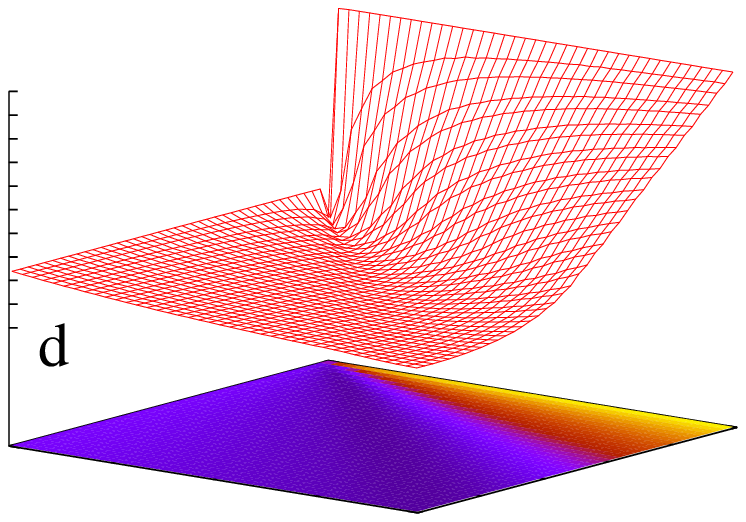}
  \includegraphics[angle=0,bb=34 38 248 188,clip,width=0.49\columnwidth]{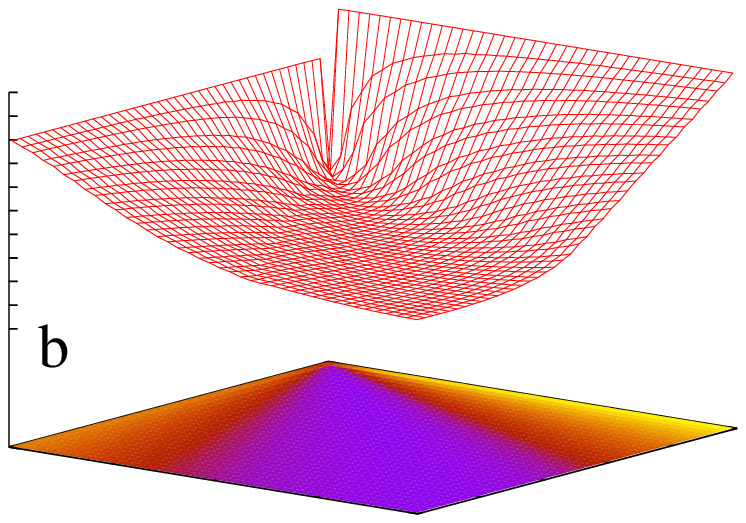}\includegraphics[angle=0,bb=34 38 248 188,clip,width=0.49\columnwidth]{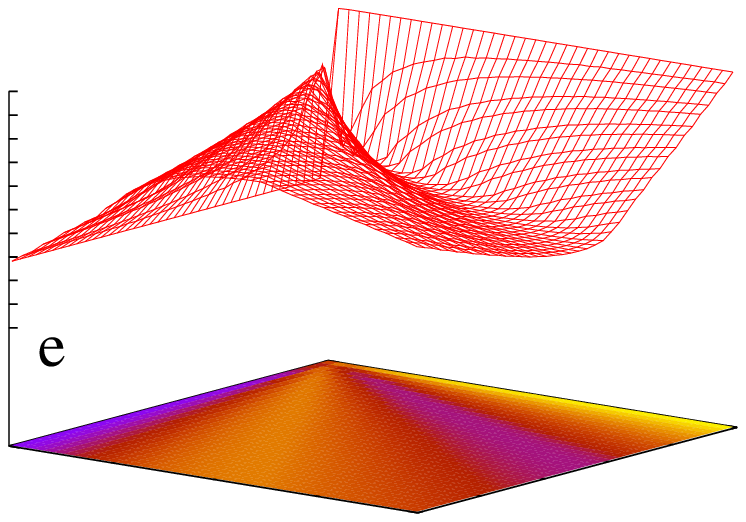}
  \includegraphics[angle=0,bb=34 38 248 188,clip,width=0.49\columnwidth]{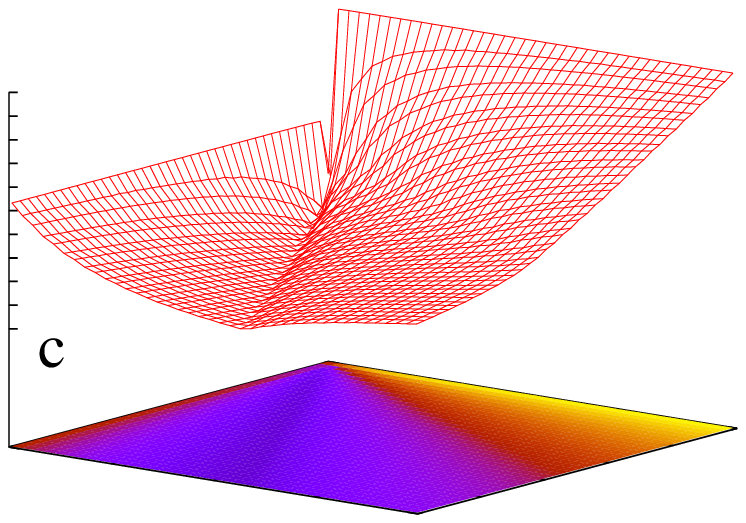}\includegraphics[angle=0,bb=34 38 248 188,clip,width=0.49\columnwidth]{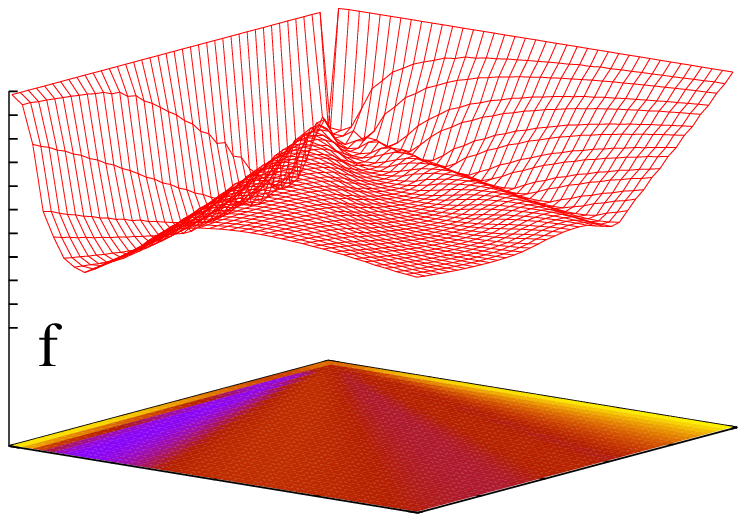} 
  \caption{The coefficient of tangential restitution as function of the normal and the tangential components of the impact velocity. 
The top figure shows the case of pure tangential damping, that is, $\varepsilon_n=1$. The spring constants acting in normal and tangential direction are $k_n=k_t=10^6$\,N/m. The left panel shows the effect of the coefficient of normal restitution on $\varepsilon_t$: a) $\varepsilon_n=0.8$, b) $\varepsilon_n=0.6$, c) $\varepsilon_n=0.4$. The right panel shows the influence of the tangential spring constant: d) $k_t=0.5\,k_n$, e) $k_t=2\,k_n$, f) $k_t=10\,k_n$. In all plots the Coulomb friction parameter is $\mu=0.4$. The numerical solution agrees perfectly with the analytical solution.}
  \label{fig:ps_krake}
\end{figure}

Before discussing more general properties of the collision model let us look to the typical structure of the coefficient of tangential restitution as it follows from the model by Cundall and Strack. The set of equations (\ref{eq:eq_mot_CS}, \ref{eq:eq_mot_tan_CS}) can be integrated numerically. Figure \ref{fig:ps_krake} shows the coefficient of tangential restitution as a function of the normal and tangential components of the impact velocity.

As a reference the plot on the top of Fig. \ref{fig:ps_krake} shows the case of elastic normal force $\gamma_n=0$. In the left panel we introduce damping of the normal force of different magnitude, i.e., $\gamma_n>0$ or $\varepsilon_n>0$, respectively. In the right panel the spring constant $k_t$ is varied. For small tangential velocity, $g_t$, the coefficient of tangential restitution approaches in all cases a constant whose value depends on both $k_t$ and $\varepsilon_n$. Contrary, in the case of large tangential velocity or small normal velocity, $\varepsilon_t(g_n,g_t)$ behaves as for the case of pure Coulomb friction. Consequently, for vanishing normal velocity, $g_n\to 0$, see Eq. \eqref{eq:coulomb_eps}.

The functional form of $\varepsilon_t(g_n,g_t)$ reveals a more complicated behavior for increasing stiffness of the tangential spring, $k_t$, see Fig \ref{fig:ps_krake} e and f. For larger stiffness the function $\varepsilon_t(g_n,g_t)$ develops an increasing number of valleys. We will discuss this behavior below in Sec. \ref{sec:gesamt}. 

In contrast, the choice of the damping in normal direction does not lead to qualitative changes of $\varepsilon_t(g_n,g_t)$. Only for very small coefficients of normal restitution we find a significant change of the form of $\varepsilon_t$ as compared to the case of pure tangential damping, $\varepsilon_n=1$.  

\subsubsection{Elastic normal spring}

The numerical results presented in the previous section suggest that the choice of the damping parameter, $\gamma_n$ or $\varepsilon_n$, in normal direction is of much less importance than the choice of the tangential spring constant $k_t$. It is worth to discuss an approximate theory for the limit of elastic interaction in normal direction, that is, $\gamma_n=0$ in Eq. \eqref{eq:dashpot}.

For given elastic and dissipative material constants, $k_n$, $k_t$, $\gamma_n=0$, $k_t$, and $\mu$, the time dependence of the compression, $\xi(t)$, and the normal velocity, $g_n(t)$, depend only on the normal component of the impact velocity at the point of contact, $g_n=g_n(0)$, but they are independent of the tangential component $g_t$. Integrating Eq. \eqref{eq:eq_mot_CS} we obtain
\begin{equation} 
  \label{eq:xioft}
  \xi(t)=\frac{g_n}{\omega_n} \sin\left(\omega_n t\right)\,,~~~~t\in\left(0,t_c\right)\,,~~~t_c=\pi/\omega\,,
\end{equation}
with $\omega_n$ defined in Eq. \eqref{eq:omegadef}.

Assume at time $t_0$ ($t_0$ is not necessarily the time of impact) the tangential component of the relative velocity is $g_t\left(t_0\right)$ and the tangential motion occurs in the Cundall-Strack regime. At this instant the tangential velocity is $g_t(t_0)$ and the elongation of the Cundall struck spring $\zeta(t_0)$. The solution of the equation of motion, Eq. \eqref{eq:eq_mot}, for the tangential velocity and the elongation of the tangential spring is then
\begin{equation} 
  \label{eq:sol_cunstrackz}
  \zeta(t) = \zeta\left(t_0\right) \cos\left[ \omega_t (t-t_0) \right] 
  +  \frac{g_t\left(t_0\right)}{\omega_t}  \sin\left[ \omega_t (t-t_0)\right] 
\end{equation}
\begin{equation}
  \label{eq:sol_cunstrackg}
  g_t(t) = -\zeta\left(t_0\right) \omega_t \sin \left[ \omega_t \left(t-t_0\right) \right]+g_t\left(t_0\right)
  \cos\left[\omega_t \left(t-t_0\right)\right],
\end{equation}
with $\omega_t=\sqrt{k_t/\alpha}$. On the other hand, if the tangential motion is governed by Coulomb's friction law, the solution of the equation of motion, Eq. \eqref{eq:eq_mot_tan_CS}, is
\begin{eqnarray}
 g_t(t)&=& g_t(t_0)-\frac{\mu}{\alpha} \text{sgn}\left[\zeta\left(t_0\right)\right] \int_{t_0}^t\,\text{d}t^\prime F_n\left(t^\prime\right) \nonumber \\   
 &=& g_t(t_0) + \text{sgn}\left[\zeta\left(t_0\right)\right] \frac{\mu g_n m_\text{eff}}{\alpha}\nonumber \\
&& ~~~~\times\left[\cos\left(\omega_n t\right)-\cos\left(\omega_n t_0\right) \right] 
 \label{eq:sol_Coulomb}
\end{eqnarray}
and the absolute value of the spring length is
\begin{align} 
  \label{eq:Co_zeta}
  \zeta(t)=\frac{\mu k_n}{k_t} \xi(t) 
\end{align}

As in the case of the Haff-Werner law, discussed in Sec. \ref{sec:Haff_analyse} the force may change during a collision between the Cundall-Strack regime, $\left|F_t\right|=k_t \zeta$, employing the elastic spring of stiffness $k_t$ and the Coulomb regime, $F_t=\mu k_n \xi$.  (For a detailed analysis of the switching properties see Sec. \ref{sec:switching}.) In order to determine the coefficient of tangential restitution we have to determine the times when the regime changes and combine the corresponding partial solutions, Eqs. \eqref{eq:sol_cunstrackg} and \eqref{eq:sol_Coulomb}, correspondingly. Whereas in case of the Haff-Werner force there are only zero or two changes of the regime, we will see that in case of the Cundall-Strack force there may occur multiple changes of the regime, see Sec. \ref{sec:switching}. 

If the motion is governed by the Cundall-Strack force, the system changes to the Coulomb regime at time $t_s$ when the Coulomb force equals the force according to the Cundall-Strack force, 
\begin{equation} 
  \label{eq:switch_CS}
  \mu k_n \xi\left(t_s\right) = k_t \left|\zeta\left(t_s\right)\right| 
\end{equation}

Contrary, if at present time the motion is governed by the Coulomb force, determining the time when the Cundall-Strack regime will take over is less straight-foreward: If the system is in the Coulomb regime, the elongation of the tangential spring is determined by Eq. \eqref{eq:Co_zeta}. Thus, in this regime the Coulomb force and the Cundall-Strack force are equal. The regime changes if after an infinitesimal time $t+dt$, the Cundall-Strack force according to Eq. \eqref{eq:CundallSpring} exceeds $\mu$ times the Coulomb force, Eq. \eqref{eq:xioft}, that is, the next switching time is determined by the time when the derivatives of the forces in both regimes equal one another. Therefore, the collision switches from the Coulomb regime to the Cundall-Strack regime at time $t_s$ with  
\begin{equation} 
  \label{eq:switch_Co}
  \frac{\mu g_n k_n}{k_t} \cos\left(\omega_n t_s\right) = g_t\left(t_s\right) 
\end{equation}  
where $g_t(t)$ is governed by Eq. \eqref{eq:sol_Coulomb}. 

Finally, we have to determine whether the collision starts in the Coulomb regime or in the Cundall-Strack regime. We consider the potential change of the Cundall-Strack force and the Coulomb force in an infinitesimal time interval. The condition
to start in the Coulomb regime reads, thus,
\begin{equation} 
  \label{eq:startcoulomb}
  \mu k_n g_n < k_t g_t 
\end{equation} 
This inequality, together with Eqs. (\ref{eq:sol_cunstrackz}-\ref{eq:sol_Coulomb}, \ref{eq:switch_CS}, \ref{eq:switch_Co})
describe the tangential motion of the colliding spheres.
We determined the coefficient of tangential restitution by combining the piecewise solutions for the Cundall-Strack regime and the Coulomb regime with regard to the corresponding switching times $t_s$. The result agrees perfectly with the numerical solution shown in Fig. \ref{fig:ps_krake}.

Let us now discuss the special case of pure Coulomb friction. If the 
condition \eqref{eq:startcoulomb} 
if fulfilled, the dynamics starts in the Coulomb 
regime and switches to the Cundall Strack regime if 
Eq. \eqref{eq:switch_Co} is fulfilled.
However there may be no real solution
of Eq. \eqref{eq:switch_Co} and hence the Coulomb regime may be active 
during the entire collision. In that case the 
coefficient of restitution is described by equation
\eqref{eq:coulomb_eps}. Using Eq. \eqref{eq:sol_Coulomb} with $t_0=0$
and inserting this in Eq. \eqref{eq:switch_Co} one finds
\begin{equation}
\frac{\mu g_n k_n}{k_t} \cos \left( \omega_n t_s \right) =
g_t+\frac{\mu g_n m_\text{eff}}{\alpha} \left[ \cos \left(\omega_n t_s\right) -1
\right]\,,
\end{equation}
assuming (without loss of generality) a positive $g_t$ yielding $\text{sgn}{\zeta}=1$. 
Solving this equation for $\cos(\omega_n t_s)$ one finds
\begin{equation}
  \mu \left(\frac{k_n}{k_t} - \frac{m_\text{eff}}{\alpha}\right)\cos (\omega_n t_s) = \frac{g_t}{g_n} - \mu\frac{m_\text{eff}}{\alpha}\,.
\end{equation} 
This equation has no real solution if
\begin{equation} \label{eq:stay_coulomb}
\left| \frac{g_t}{g_n} - \mu \frac{m_\text{eff}}{\alpha} \right| > \mu
\left| \frac{k_n}{k_t} - \frac{m_\text{eff}}{\alpha} \right| \,,
\end{equation}
as $|\cos\omega_nt_s|$ cannot be larger than one. 
Hence if both inequalities \eqref{eq:startcoulomb} and 
\eqref{eq:stay_coulomb} are fulfilled the dynamics stays in the
Coulomb regime.
Furthermore is  possible to calculate the value of $\varepsilon_t$ 
which is reached when $g_t$ tends to zero. 
It is obvious from the inequality \eqref{eq:startcoulomb}
that the  dynamic starts for small $g_t$ in the Cundall-Strack
regime and that there is a small time interval before the end of the
collision where the dynamics is governed by the Coulomb regime. This time interval is proportional to $g_t/g_n$ which is small by construction (cf. Eq. \eqref{eq:switch_CS} together with Eqs. \eqref{eq:xioft} and \eqref{eq:sol_cunstrackz}). Hence, the expression $\cos(\omega_nt_0)$ in Eq. \eqref{eq:sol_Coulomb} is $\cos(\omega_nt_c) + {\cal O}((g_t/g_n)^2)$. The small correction ${\cal O}((g_t/g_n)^2)$ can be neglected and the final velocity $g_t^\prime$ can be approximated as $g_t^\prime=g_t\cos(\omega_tt_c)$ yielding
\begin{equation} \label{eq:van_tan_vel}
  \lim_{g_t \rightarrow 0} \varepsilon_t(g_n,g_t) = \cos(\omega_t t_c) =
  \cos\left(\frac{\pi \omega_t}{\omega_n} \right) \,,
\end{equation}
(cf. Eq. \eqref{eq:sol_cunstrackg} with $t_0$=0 and $t=t_c$). 
  
\subsubsection{Scaling properties for the case of elastic normal springs}
\label{sec:scaling}
   
\begin{figure}
  \centering
  \ifthenelse{\equal{\ColBW}{color}}{
  \includegraphics[bb=27 30 396 258,clip,width=0.99\columnwidth]{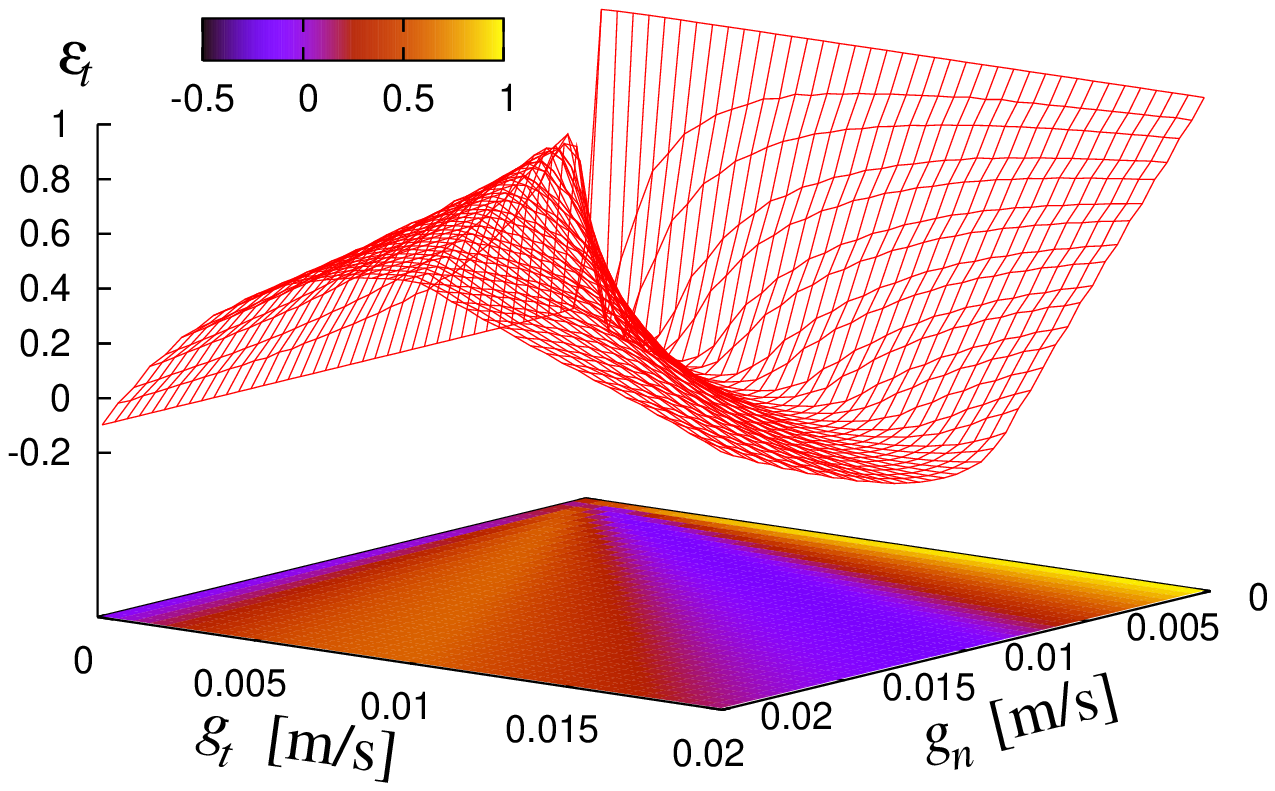}
  \includegraphics[bb=27 30 396 258,clip,width=0.99\columnwidth]{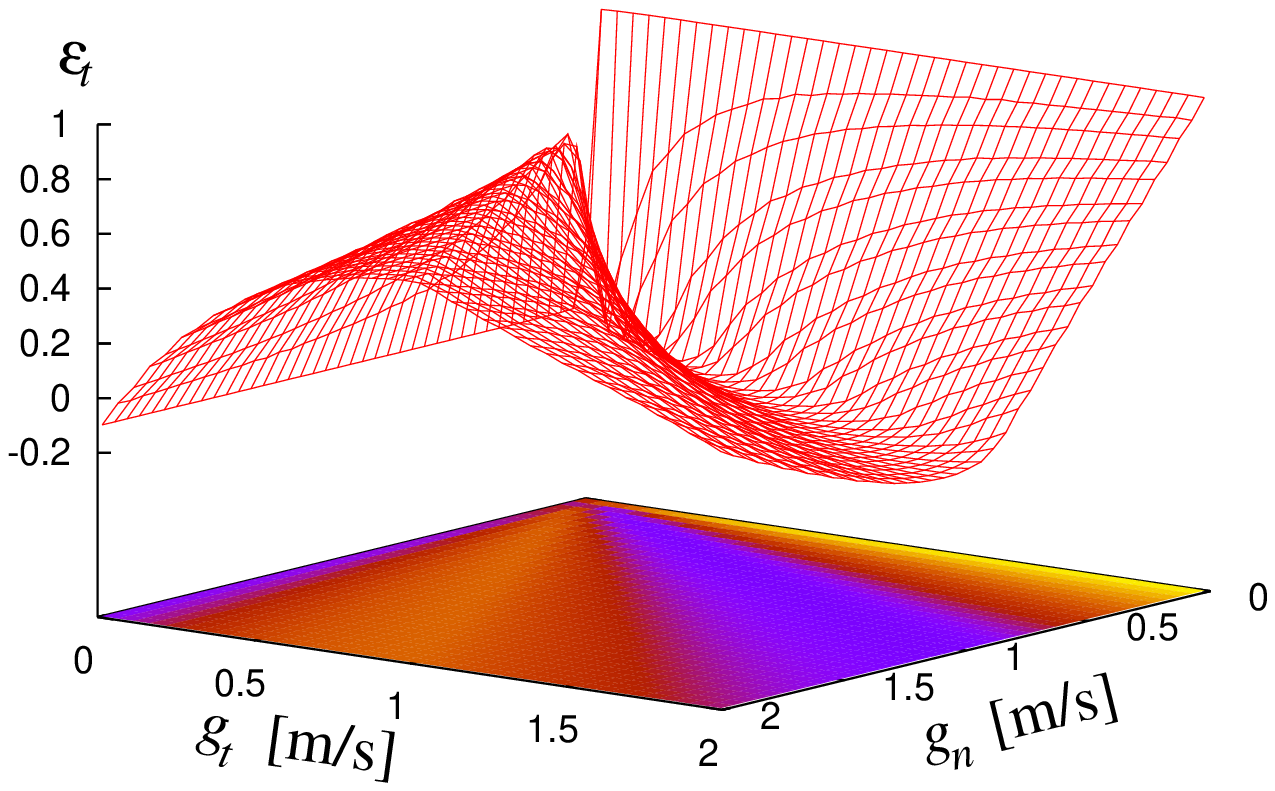}}
{
  \includegraphics[bb=27 30 396 258,clip,width=0.99\columnwidth]{figs/scaling/scaling2.eps}
  \includegraphics[bb=27 30 396
  258,clip,width=0.99\columnwidth]{figs/scaling/scaling1.eps}}

  \caption{For elastic normal restitution, the Cundall-Strack model reveals the simple scaling form, Eq. \eqref{eq:CSscalingEOM}. Despite of its rather complicated functional form, the plot of $\varepsilon_t\left(g_n,g_t\right)$, is invariant when scaling the impact velocity. Top: velocity range $0\ldots 0.02$ m/sec, bottom $0\ldots 2$ m/sec. The other parameters are $k_n=k_t=10^6$ N/m, $\mu=0.4$. The pictures are identical since the coefficient of restitution depends only on $g_n/g_t$.}
  \label{fig:eps_scaling}
\end{figure}

For the case $\gamma_n=0$, that is, when the dissipation of the motion in normal direction can be neglected, $\varepsilon_n=1$, apparently, the collision as described by the Cundall-Strack model depends on 7 parameters: $m$, $J$, $\mu$, $k_n$, $k_t$, $g_n$, and $g_t$. By using appropriate time and length scales we can reduce the number of free parameters to 3. The length scale is the maximum compression $\xi_\text{max}$ in normal direction:
\begin{equation}
  \xi_\text{max} = \sqrt{\frac{m_\text{eff}}{k_n}}g_n = \frac{g_n}{\omega_0}
\end{equation}
The obvious time scale $T$ of the problem is the duration of the collision, $t_c=\pi/\omega_0$. To simplify the resulting expressions we drop the prefactor $\pi$ and define
\begin{equation}
  T \equiv \frac{1}{\omega_0} = \sqrt{\frac{m_\text{eff}}{k_n}}\,.
\end{equation}
The scaled variables are, thus, the scaled deformation, the scaled length of the tangential spring, and the scaled time, 
\begin{equation}
  \label{eq:CSscaling}
  x\equiv\xi/\xi_\text{max}\,;~~~z\equiv\zeta/\xi_\text{max}\,;~~~\tau\equiv t/T\,.  
\end{equation}
Taking into account that $\alpha/m_\text{eff}$ with $\alpha$ defined
by Eq. \eqref{eq:alphadef}, for identical homogeneous spheres reduces to a pure number, $\alpha/m_\text{eff}=2/7$, the equation of motion, Eq. \eqref{eq:eq_mot_CS} is
\begin{equation}
  \label{eq:CSscalingEOM}
  \begin{split}
  & \ddot{x} + x = 0\\
  & \ddot{z} + \frac{7}{2}\mbox{sgn}\zeta\min\left(\frac{k_t}{k_n}z, \mu x\right) = 0 \\
  & x(0) = 0\,;~~ z(0) = 0\,;~~\dot{x}(0) = 1\,;~~\dot{z}(0) = \frac{g_t}{g_n}\,,
  \end{split}
\end{equation}
where dots denote time derivatives with respect to the scaled time $\tau$.
Consequently, the parameters of the system are $\mu$, $k_t/k_n$, and $g_t/g_n$. As an example, the dependence on $g_t/g_n$ is demonstrated in Fig. \ref{fig:eps_scaling}. When scaling both velocities by 100, the resulting picture is identical. 

\subsubsection{Switching between friction regimes}
\label{sec:switching}

During a collision, depending on the impact velocity and the material parameters the relative motion of the particles at the point of contact may change its character, namely it may change to and fro the Coulomb regime where the friction force is determined (i.e. capped) by the normal force and the Cundall-Strack regime where the magnitude of the friction force is determined by the length of the tangential spring alone. 

This change of regime occurs also in the Haff-Werner model, however, there are fundamental differences: As shown in Sec. \ref{sec:Haff_analyse} there may be only zero or two changes, for the Cundall-Strack model we may have multiple changes. This property originates from the fact that in the Cundall-Strack regime there is no loss of energy. Instead, during the Cundall-Strack regime the energy of the relative motion is used to load the tangential spring whose energy can be released subsequently, that is, the tangential spring acts as a reservoir of energy. The only way to dissipate the energy stored in the tangential spring is by switching into the Coulomb regime and rapidly decreasing the elongation of the tangential spring due to decreasing normal force. In this case the energy in the spring cannot be fully recovered. 
This tangential spring is a great advantage of the Cundall-Strack model as its action may lead to a negative coefficient of tangential restitution which cannot be achieved by the Haff-Werner model.

The number of switching events between the regimes as described by the criteria, Eqs. \eqref{eq:switch_CS} and \eqref{eq:switch_Co} and the initial regime given by Eq. \eqref{eq:startcoulomb} is shown in Fig. \ref{fig:postimeev}. 
\begin{figure}[htbp]
  \centering
  \ifthenelse{\equal{\ColBW}{color}}{
    \includegraphics[bb=13 13 669 546,clip,width=0.99\columnwidth]{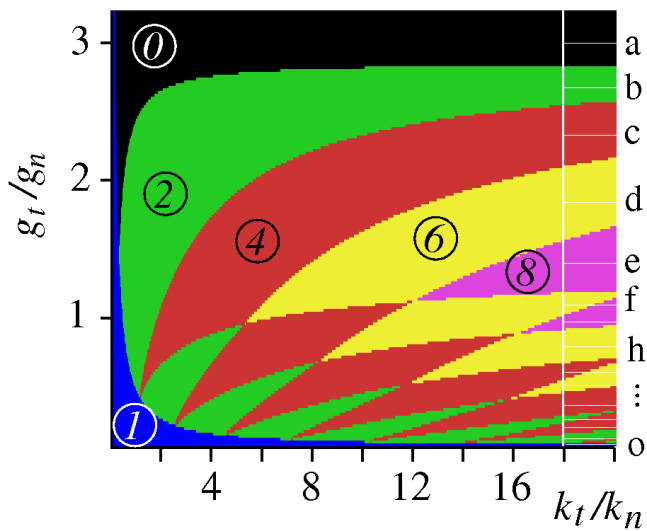}}{
    \includegraphics[bb=0 -1 659 531,clip,width=0.99\columnwidth]{figs/schaltstellen/schalterBetterBW.eps}
  }
  \caption{Number of switching events between the Cundall-Strack and the Coulomb regimes during a collision as a function of $g_t/g_n$ and $k_t/k_n$. The numbers of changes are indicated by gray shading (color online) and by the numbers in circles. The Coulomb friction coefficient is $\mu=0.4$. The letters $a,\ldots,o$ at the right side (not all are shown) refer to qualitatively different functions $\zeta(t)$ or $z(\tau)$, shown in Fig. \ref{fig:feder}.}
  \label{fig:schaltpunkte}
\end{figure}

For any value of $k_t/k_n$ there is a critical $g_t/g_n$ above which
the particle stays in the Coulomb regime (region ``0'' -- black color
in Fig. \ref{fig:schaltpunkte}), corresponding to case of pure Coulomb
friction discussed in Sec. \ref{sec:PureCoulomb}. The boundary of this
region is given by Eq. \eqref{eq:startcoulomb} and Eq. \eqref{eq:stay_coulomb}

\begin{figure}
  \centering
\includegraphics[width=0.99\columnwidth,clip]{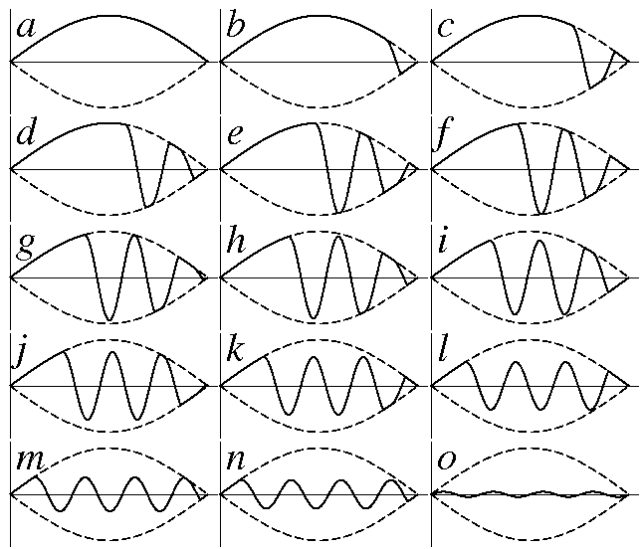}
  \caption{Elongation of the cundall Strack Spring as funtion of time for $k_t/k_n=18$ and the values $g_t/g_n$ indicated by ``a'' to ``o'' in Fig. \ref{fig:schaltpunkte}. The Couloumb limit is shown by dashed lines.}
  \label{fig:feder}
\end{figure}

Decreasing the tangential component of the impact velocity below the limit of pure Coulomb friction the collision will switch to the Cundall-Strack regime, at least for a short period of time. It may repeatedly switch back and forth between the two regimes, dependent on the impact velocity and the spring constants $k_n$ and $k_t$. To demonstrate this effect we marked 15 points in Fig. \ref{fig:schaltpunkte} indicated by ``a'' to ``o'' at the right-hand side of the figure (not all are shown) which correspond to decreasing $g_t/g_n$ for fixed $k_t/k_n$. For these parameter combinations we show in Fig. \ref{fig:feder} the elongation of the tangential spring over time (full lines). 
The dashed lines show the Coulomb limit, i.e. $\mu F_N/k_t$. Indeed, for large enough tangential velocity (Fig. \ref{fig:feder}a) the tangential force remains on the Coulomb limit for the entire duration of the collision. 

For tangential velocities slightly smaller than the limit given by Eq. \eqref{eq:startcoulomb} (region ``b'' in Fig. \ref{fig:schaltpunkte} and Fig. \ref{fig:feder}b) the collision switches to the Cundall-Strack regime near its end, that is, the friction force is sufficient to stop the particle once. For the chosen value of the spring constants ($k_t\gg k_n$) the period of the tangential oscillation is much smaller than the period of the normal oscillation. Therefore, the tangential relative velocity of the particles at the point of contact is reverted. Finally, the Coulomb limit takes effect again, the particle stays in the Coulomb regime to the end of the collision. As the tangential motion is reverted once the coefficient of tangential restitution is negative. 

For still a little smaller velocity (region ``c'' in Fig. \ref{fig:schaltpunkte} and Fig. \ref{fig:feder}c) the collision switches earlier from the Coulomb regime to the Cundall-Strack regime as its smaller tangential energy is dissipated earlier. Hence, the remaining time of contact is large enough to allow not only the reversal of motion and the subsequent switch back to the Coulomb regime as in the case ``b'' but allows an additional switch back to the the Cundall-Strack regime, that is,  the tangential velocity changes its sign back to the original direction. Finally the collision switches back to the Coulomb regime. Since the tangential velocity has the same sign as at the time of the impact, the coefficient of tangential restitution is positive. 

Note that the amplitude of the first tangential oscillation (after the first switch to the Cundall-Strack regime) {\em increases} with decreasing tangential velocity -- as the time of the first switch shifts towards the time of maximal compression. Therefore, the oscillation cannot complete even a half period without transiting to the Coulomb regime. Therefore, the number of switches is twice the number of zeros in the elongation of the Cundall-Strack spring. 

For further decreasing tangential velocity this mechanism is repeated, the number of switches increases in steps of two. After the maximum number of switches (8) is achieved for the region ``e'' in Fig. \ref{fig:schaltpunkte} and Fig. \ref{fig:feder}e the number of switches decreases. This is due to the fact that the amplitude of the tangential oscillation now decreases with decreasing tangential velocity since the first switch to the Cundall-Strack regime takes place before the point of maximal compression. The number of switches now decreases as the first oscillation of the Cundall-Strack spring may now complete more than half a period. Interestingly, there is a short interval of $g_t$ where the number of switches is back to 8 due to an additional pair of switches close to the end of the collision. 

For small values of the tangential velocity or small $k_t$ there is only one switch. The particle starts in the Cundall-Strack regime and performs several tangential oscillations without violating the Coulomb condition. Only at the very end of the collision the particle switches to the Coulomb regime. As mentioned before, this is the only time when energy of the tangential motion is actually dissipated.

\subsubsection{Commensurable spring constants}

\begin{figure}
  \centering
  \includegraphics[clip,angle=0,width=0.8\columnwidth]{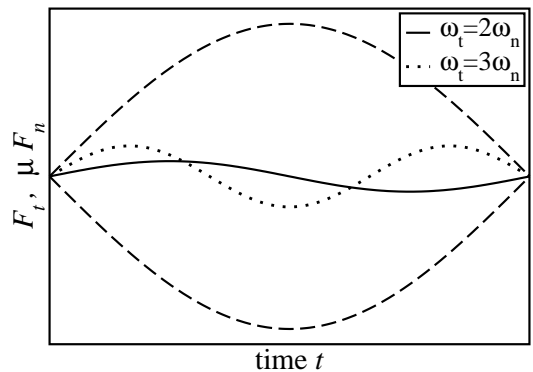}
  \caption{Sketch of the elongation of the tangential spring for commensurable frequencies, $\omega_t=2 \omega_n$ and $\omega_t=3 \omega_n$, for the case that the collision starts in the Cundall-Strack regime.}
  \label{fig:plateauskizze}
\end{figure}

Let us discuss briefly the special case of commensurable tangential and normal motion that occurs when the elastic constants $k_n$ and $k_t$ are such that the frequency of the tangential Cundall-Strack spring is a multiple of frequency of the normal motion, $\omega_t=m\, \omega_n$, $m=1,2,3\ldots$, as sketched in Fig. \ref{fig:plateauskizze}.

\begin{figure}
  \centering
  \ifthenelse{\equal{\ColBW}{color}}{
  \includegraphics[bb=24 33 395 255,clip,width=0.99\columnwidth]{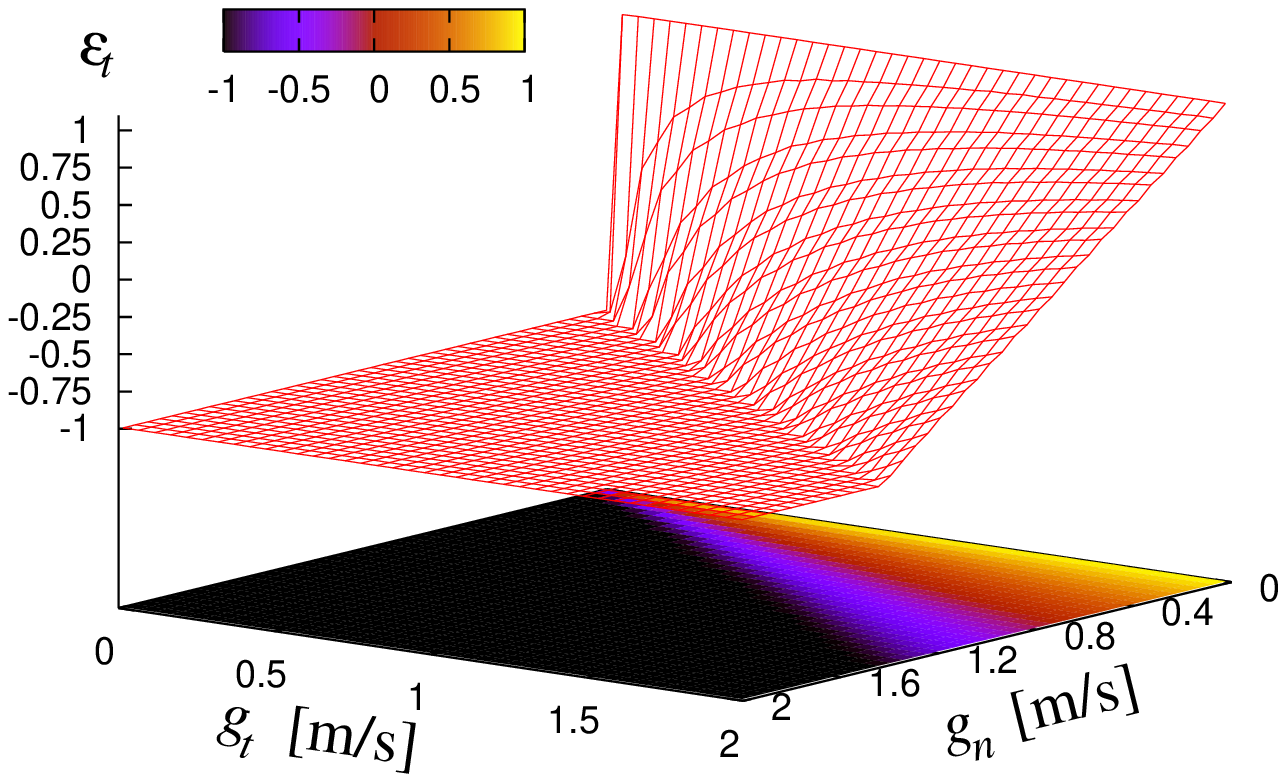}
  \includegraphics[bb=24 33 395 255,clip,width=0.99\columnwidth]{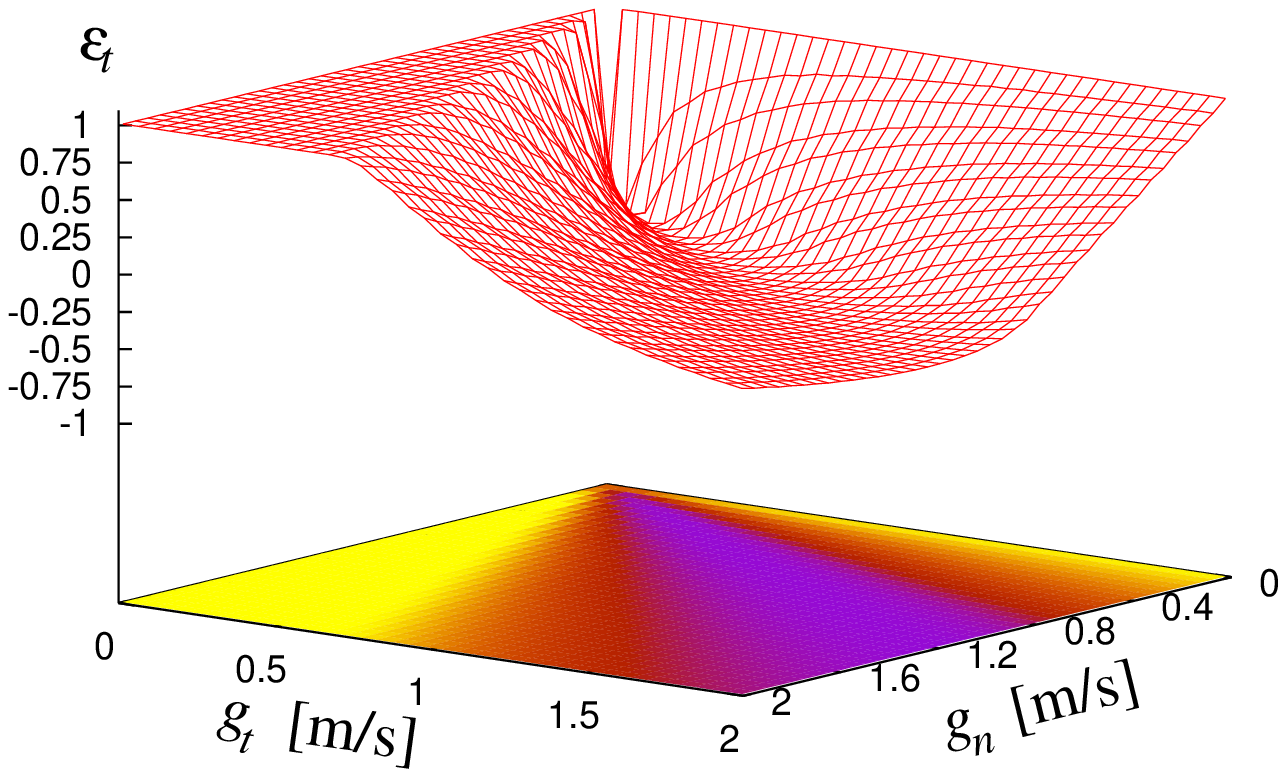}
  \includegraphics[bb=24 33 395 255,clip,width=0.99\columnwidth]{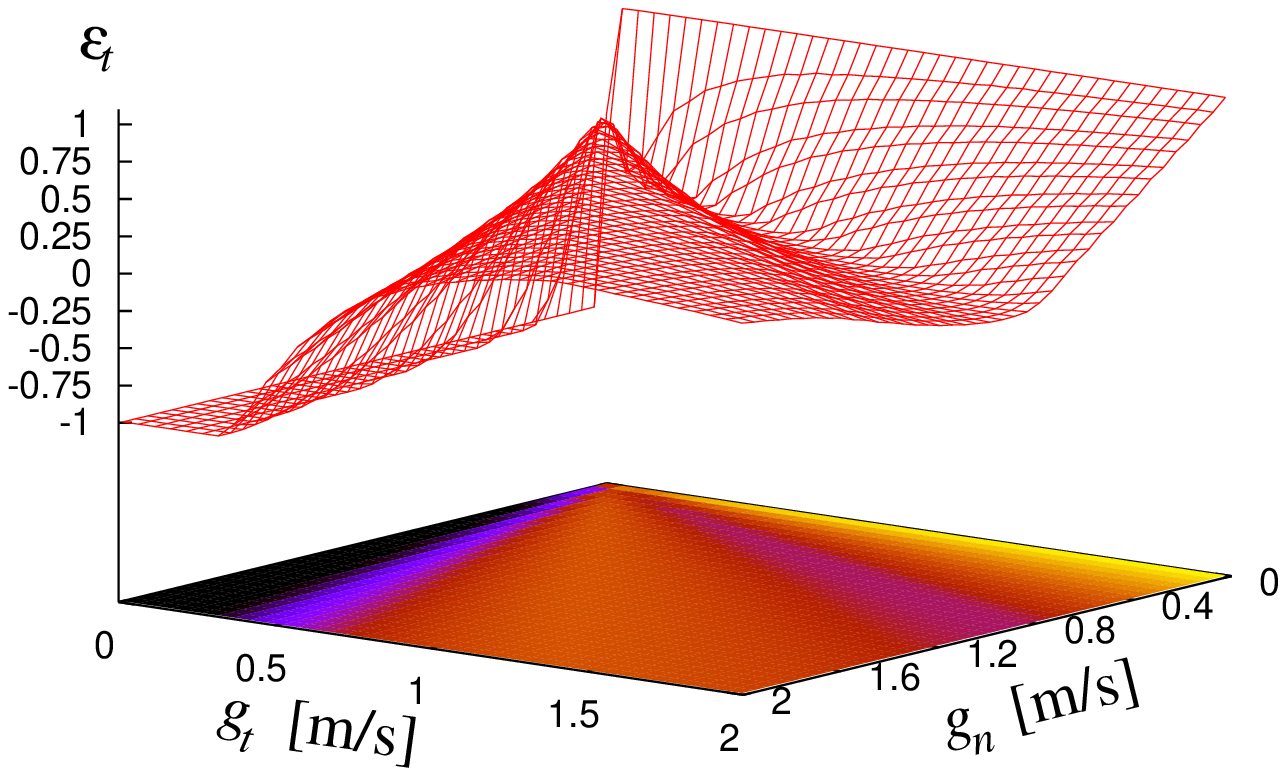}}
{ \includegraphics[bb=24 33 395 255,clip,width=0.99\columnwidth]{figs/plateaus/plateau1.eps}
  \includegraphics[bb=24 33 395 255,clip,width=0.99\columnwidth]{figs/plateaus/plateau2.eps}
  \includegraphics[bb=24 33 395 255,clip,width=0.99\columnwidth]{figs/plateaus/plateau3.eps}}
  \caption{The coefficient of tangential restitution as a function $g_n$ and $g_t$ for commensurable frequencies of the springs in normal and tangential direction. In this special case, $\varepsilon_t(g_n, g_t)$ reveals pronounced plateaus. Top: $\omega_t=\omega_n$, middle: $\omega_t=2 \omega_n$, bottom: $\omega_t=3 \omega_n$.}
  \label{fig:plateaus}
\end{figure}

In this case we notice a plateau in the coefficient of tangential restitution as a function of the components of the impact velocity, Fig. \ref{fig:plateaus}.
 
This behavior becomes clear from the sketch in Fig. \ref{fig:plateauskizze}:
If the components of he impact velocity, $g_n$ and $g_t$ are such that the collision starts in the Cundall-Strack regime and $\omega_t$ is a multiple of $\omega_n$, the elongation of the tangential spring is zero at the end of the collision. 
Hence, in contrast to non-commensurable frequencies, the system does not transit into the Coulomb regime close to the end of the collision, that is, the entire collision takes place in the Cundall-Strack regime. As explained above, energy is only dissipated in the Coulomb regime, consequently the coefficient of tangential restitution is $\varepsilon_t=\pm 1$. Whether the value is 1 or -1 depends only on the ratio between the frequencies, 
\begin{equation}
  \varepsilon_t=(-1)^{m}\,.
\end{equation}

The condition for the collision to start in the Cundall-Strack regime is described by the inequality, Eq. \eqref{eq:startcoulomb}, therefore the boundary of the plateau is given by
\begin{equation}
  g_t=\frac{\mu k_n}{k_t} g_n\,.
\end{equation}

\subsubsection{Coefficient of tangential restitution in scaled units}
\label{sec:gesamt}

\begin{figure}
  \centering
  \includegraphics[bb=0 0 508 421,clip,width=0.99\columnwidth]{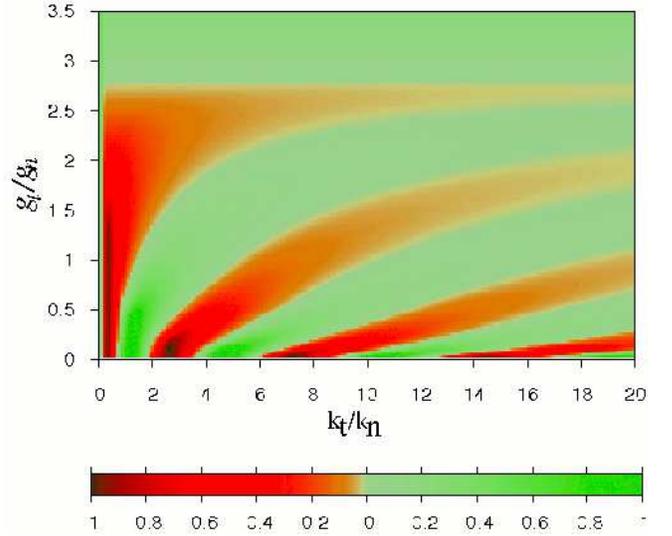}
  \caption{The coefficient of tangential restitution as a function of the ratios $g_t/g_n$ and $k_t/k_n$. The Coulomb coefficient is $\mu=0.4$.}
  \label{fig:ps_gesamt}
\end{figure}

As elaborated in Sec. \ref{sec:scaling}, the coefficient of tangential restitution does not explicitly depend on the parameters $g_t$, $g_n$, $k_n$, and $k_t$ but only on the ratios $k_t/k_n$ and $g_t/g_n$. This scaling property allows to present the coefficient of tangential restitution in a more general way it was shown Fig. \ref{fig:ps_krake}. Figure \ref{fig:ps_gesamt} shows $\varepsilon_t\left(k_t/k_n, g_t/g_n\right)$. 

For sufficient large tangential velocity we recover the pure Coulomb regime (top region in Fig. \ref{fig:ps_gesamt}. For very small $g_t/g_n$, $\varepsilon_t$ oscillates between $-1$ and $1$. As explained above, in the limit of vanishing tangential velocity, $g_t$, the coefficient of tangential restitution is described by Eq. \eqref{eq:van_tan_vel}. Expressed in terms of the ration $k_t/k_n$ one obtains
\begin{equation}
  \varepsilon_t\left( \frac{g_t}{g_n} \rightarrow 0 ,
  \frac{k_t}{k_n}\right) = \cos \left(\pi \sqrt{\frac{k_t}{k_n}\frac{m_\text{eff}}{\alpha}  } \right)\,.
\end{equation}

The oscillating behavior of $\varepsilon_t\left(g_t/g_n, k_t/k_n\right)$ in both directions for fixed $g_t/g_n$ and varying $ k_t/k_n$ as well as for fixed $ k_t/k_n$ and varying $g_t/g_n$ may be attributed to the switching between the Coulomb regime and the Cundall-Strack regime as discussed in Sec. \ref{sec:switching} (see Figs. \ref{fig:schaltpunkte} and \ref{fig:feder}).

Note that the green line in the very left of Fig. \ref{fig:ps_gesamt} is not an artifact of plotting. Here the coefficient of tangential restitution rises very steeply to $\varepsilon_t\to 1$ since for very small $g_t$ during the collision whose duration is determined by $k_n$ the tangential spring cannot be elongated enough to transit into the Coulomb regime.

\section{Variation of the unit vector $\vec{e}$ during the collision}
\label{sec:unitvector}

\begin{figure}
  \centering
  \includegraphics[width=0.7\columnwidth,bb= 0 0 265 113]{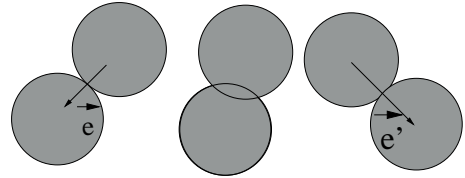}
  
  \caption{During a non-central collision the unit vector $\vec{e}\equiv \left(\vec{r}_i-\vec{r}_j\right)/\left|\vec{r}_i-\vec{r}_j\right|$ changes its direction.}
  \label{fig:unit_sketch}
\end{figure}

All results in the previous section were obtained under the assumption that the unit vector $\vec{e}\equiv \left(\vec{r}_i-\vec{r}_j\right)/\left|\vec{r}_i-\vec{r}_j\right|$ keeps constant during the entire collision. This assumption is exact only for a central collision when $g_t=0$. If the particles collide with finite tangential relative velocity, the vector $\vec{e}$ cannot be constant but changes its direction, as shown in Fig. \ref{fig:unit_sketch}.   

\begin{figure}
  \centering
  \includegraphics[angle=0,bb=31 19 391 270,clip,width=0.99\columnwidth]{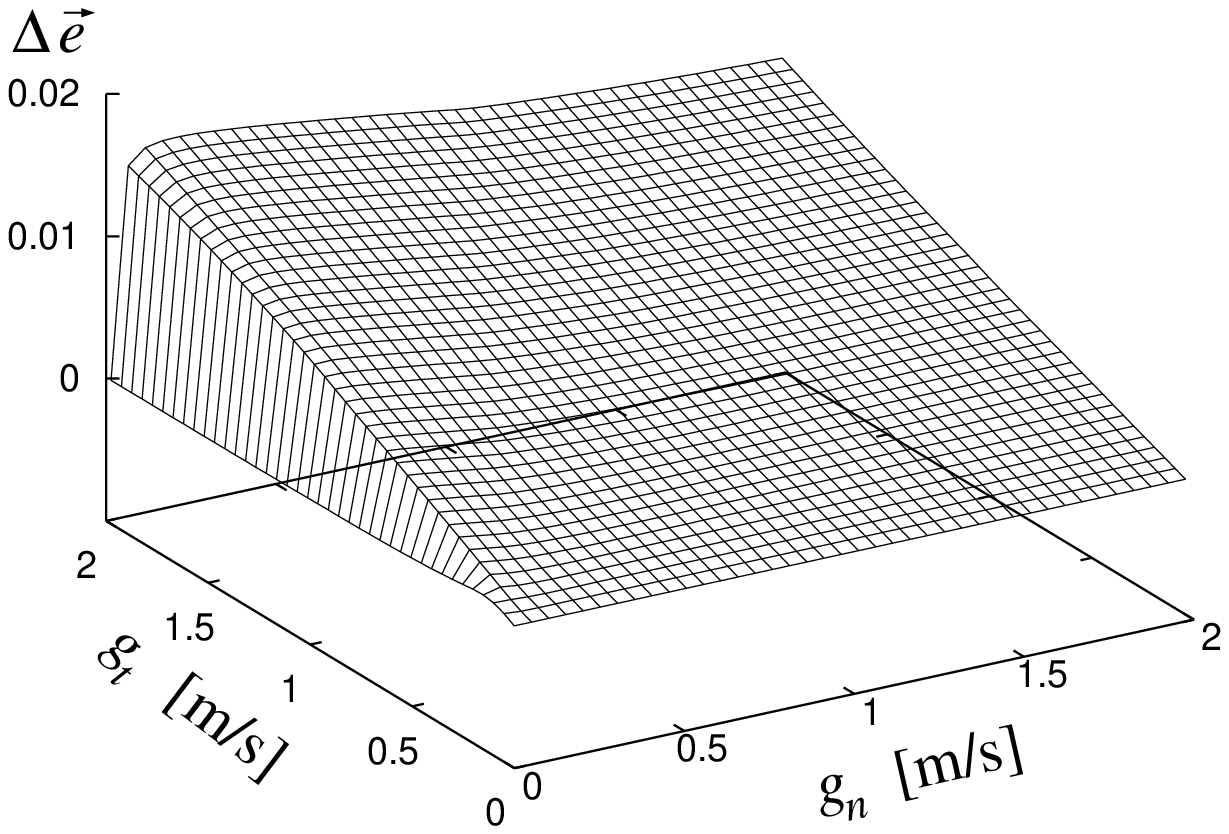}
  \includegraphics[angle=0,bb=31 19 391 270,clip,width=0.99\columnwidth]{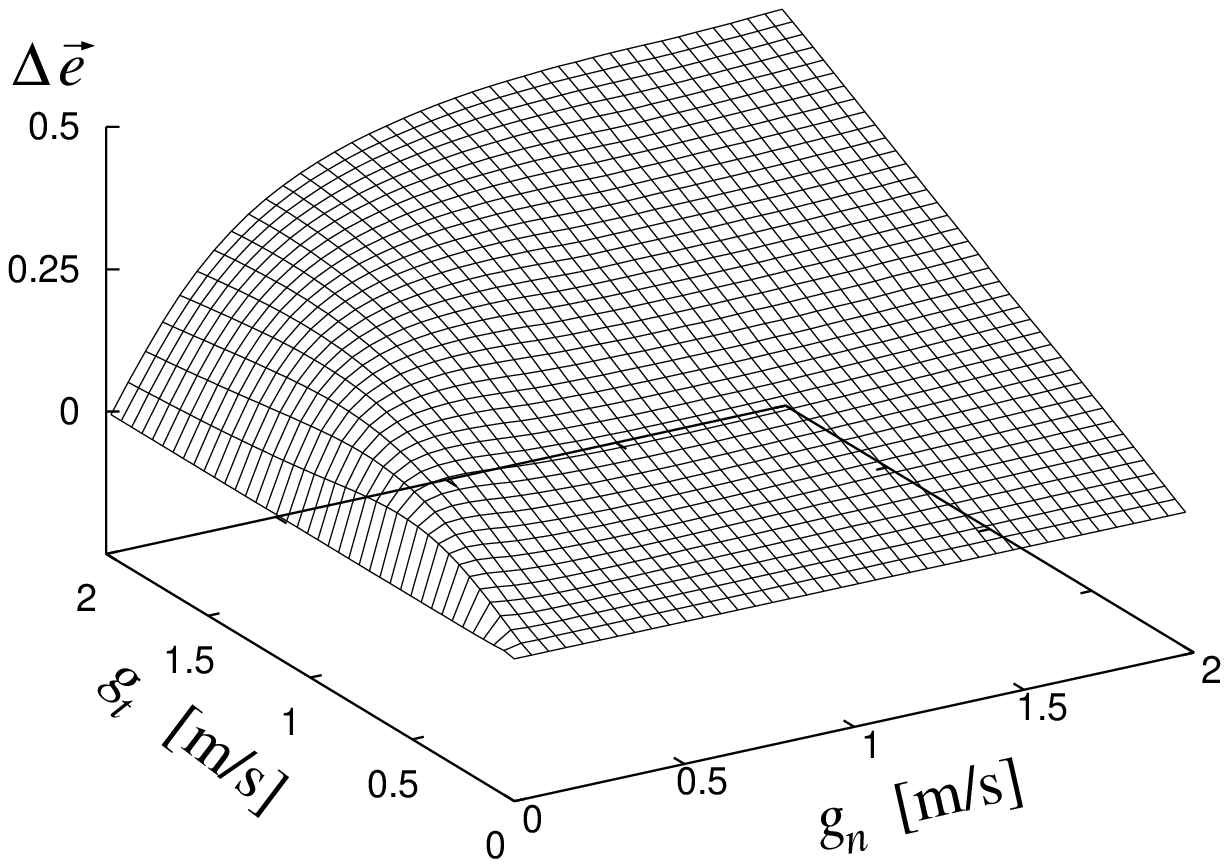}
  \caption{Variation of the unit vector
    $\vec{e}=\vec{r}_i-\vec{r}_j/\left|\vec{r}_i-\vec{r}_j\right|$
    during a collision as function of $g_n$ and $g_t$. The change
    $\Delta\vec{e}$ is defined as the angle in rad between the
    $\vec{e}$ in the beginning and the end of the collision. The
    parameters are $\mu=0.4$, $m_1=m_2=1$g, $R_1=R_2=4$~mm, 
    $k_n=k_t=10^6$~N/m (top), $k_n=k_t=10^3$~N/m (bottom). }
  \label{fig:ps_winkel}
\end{figure}

By means of numerical integration of Newton's equation of motion for a
pair of colliding spheres of mass $m_1=m_2=1$~g and radii $R_1=R_2=4$~mm. We computed the variation of the unit vector $\vec{e}$ during the collision to check whether the assumption $\vec{e}\approx$const. is justified. Figure \ref{fig:ps_winkel} 
shows the change $\Delta\vec{e}$ as a function of the normal and tangential components of the impact velocity. Here $\Delta\vec{e}$ is defined as the angle in rad between the unit vector at the beginning of the collision and its end. 

For rather soft particles, $k_n=10^3$\,N/m (bottom part of Fig \ref{fig:ps_winkel}) the unit vector changes remarkably up to about 30$^o$ whereas for more stiff particles, $k_n=10^6$\,N/m (top part) the angle is below 1$^o$. Hence, for sufficiently hard particles the assumption $\vec{e}=$const. is justified.  

\section{Conclusions}

We investigated the coefficient of tangential restitution for linear normal forces and two different tangential force models -- the models by Haff and Werner \cite{HaffWerner:1986} and by Cundall and Strack \cite{CundallStrack:1979}. 

For the model by Haff and Werner, we showed that the coefficient of restitution is strictly non-negative. In a good approximation its functional form can conveniently be described as either a (non-negative) constant or the dependence given by Eq. \eqref{eq:coulomb_eps}, whatever is larger (see Eq. \eqref{eq:epsHaffWernerAproximation}). Thus, this model is unsuitable for describing collisions with negative coefficient of tangential restitution.

For the model by Cundall and Strack the coefficient of tangential restitution shows a very complex behaviour. For certain combinations of impact velocities and material parameters one may observe a negative coefficient of tangential restitution. By adopting suitable length- and time-scales one can conveniently present the velocity dependence of $\varepsilon_t$ by only three parameters, the friction coefficient $\mu$, the ratio of the compontents of the impact velocity $g_t/g_n$ and the ratio of the tangential and normal spring $k_t/k_n$, provided the dissipation of the normal spring can be neglected ($\varepsilon_n=1$). We showed that the latter parameter $k_t/k_n$ is critical for the sign of $\varepsilon_t$. 

For the limit of sufficiently large tangential velocity there is a universal velocity dependence of $\varepsilon_t(g_n,g_t)$ which is not only independent of the tangential but also of the normal force law. We call this limit the limit of pure Coulomb force. 

\vfill
\bibliography{EpsTanLinear}
\end{document}